%% file: main.tex
\documentclass[twoside,11pt]{article}
\usepackage{amsmath,bbm,fancyhdr,a4,epsfig,umlaut,amssymb,psfrag,multirow,exscale,caption2,graphicx}
 \numberwithin{equation}{section}
\numberwithin{figure}{section}
 \allowdisplaybreaks[1]
\newcommand{\be}{\begin{equation}}
\newcommand{\ee}{\end{equation}}
\newcommand{\bea}{\begin{eqnarray}}
\newcommand{\eea}{\end{eqnarray}}
\newcommand{\ba}{\begin{align}}
\newcommand{\ea}{\end{align}}
\newcommand{\refe}[1]{Eqn.\/~(\ref{#1})}
\newcommand{\refes}[2]{Eqns.\/~(\ref{#1}--\ref{#2})}
\newcommand{\refec}[2]{Eqns.\/~(\ref{#1},\:\ref{#2})}
\newcommand{\transpose}{\mathrm{t}}
\newcommand{\stranspose}{\mathrm{st}}
\newcommand{\R}{\mathbbm{R}}
\newcommand{\C}{\mathbbm{C}}
\newcommand{\uosp}{U\!O\!Sp(1|2)}
\newcommand{\sdet}{\mathrm{sdet}}

\title{
\vskip -70pt
\begin{flushright}
{\normalsize \ DAMTP-2004-98}\\
\end{flushright}
\vskip 25pt {\bf A geometric approach to scalar field theories on
the supersphere} } \vspace{1.4cm} {\makeatletter
\author{A. F. Schunck\thanks{e-mail address: A.F.Schunck@damtp.cam.ac.uk}
\hspace{1pt} and Chris Wainwright{\thanks{e-mail address:
C.J.Wainwright@damtp.cam.ac.uk}} \\ \small{\textsl{Department of Applied
    Mathematics and Theoretical Physics}}
\\ \small{\textsl{University of Cambridge}} \\
\small{\textsl{Wilberforce Road, Cambridge CB3 0WA, England}}}
\makeatother}
\date{\today}
\begin{document}
\maketitle
\input{abs}
\input{intro}
\input{uosp}
\input{coset}
\input{isometries}
\input{superzweibein}
\input{superfields}
\input{fieldtheory}
\input{susybreaking}
\input{noether}
\input{conclusion}
\input{acknowledgements}
\input{appendix}
\input{references}
\end{document}

%% file: abs.tex
\begin{abstract}
Following a strictly geometric approach we construct globally
supersymmetric scalar field theories on the supersphere, defined
as the quotient space $S^{2|2} = UOSp(1|2)/\mathcal{U}(1)$. We
analyze the superspace geometry of the supersphere, in particular
deriving the invariant vielbein and spin connection from a
generalization of the left-invariant Maurer-Cartan form for Lie
groups. Using this information we proceed to construct a
superscalar field action on $S^{2|2}$, which can be decomposed in
terms of the component fields, yielding a supersymmetric action on
the ordinary two-sphere. We are able to derive Lagrange equations
and Noether's theorem for the superscalar field itself.

\vspace{3ex} \noindent PACS numbers: 11.30.Pb, 12.60.Jv

\end{abstract}

%% file: intro.tex
\section{Introduction}
While superspheres have been extensively studied as target spaces
for supersymmetric sigma models, see e.g.\/~\cite{spanishguy,
saleur}, little attention has been paid to considering the
supersphere as the base space for supersymmetric field theories.
However, treating the supersphere as such provides us with an
interesting model for studying globally supersymmetric field
theories in curved space.

In this paper we present a strictly geometric approach to
constructing globally supersymmetric scalar field theories on the
supersphere, defined here as the coset space
$\uosp/\mathcal{U}(1)$ \cite{thepaper, landimarmo}, the body of
which is given by the ordinary two-sphere. We should emphasize
here that there is an ambiguity in defining a supersphere,
i.e.\/~a supersymmetric generalization of the ordinary two-sphere,
the only criterion being that the body of the respective
supermanifold coincides with $S^2$. Another example of a
supersymmetric generalization of $S^2$ would be the quotient space
$SU(2|1)/U(1|1)$, as considered in e.g.\/~\cite{townsend}. If one
insists, however, on the additional condition that the resultant
coset space is not just a supermanifold but rather a
\emph{superspace}, this excludes for example the latter
possibility and leaves as one obvious choice precisely the coset
space $\uosp/\mathcal{U}(1)$.

While it is not important to insist on this additional condition
for the purpose of using the supermanifold as the target space for
some supersymmetric sigma model, it is crucial to enforce it if
one wants to construct a field theory on the supermanifold as the
background. This is because the superspace condition ensures
firstly that the tangent space group of the supermanifold under
consideration corresponds to the even Grassmann extension of the
tangent space group of the body of the respective coset space and
secondly that the fermionic field content of the theory will
transform as \emph{spinor} fields under the action of the tangent
space group (see Sections \ref{sec:othersuperspheres},
\ref{sec:vielbeinspin}).

Note, however, that taking the coset space $\uosp/\mathcal{U}(1)$
as the supersymmetric generalization of the ordinary sphere
involves inevitably the usage of a rather unfamiliar extension of
complex conjugation to supernumbers, referred to as
pseudo-conjugation \cite{scheunert}, see Section
\ref{sec:pseudo-conjugation}, together with the definition of a
graded adjoint, see Section \ref{sec:gradedadjoint}.

We shall emphasize one other important point about our approach to
constructing scalar field theories on $S^{2|2}$. While it is
possible to construct supersymmetric theories on certain curved
backgrounds using component fields from the outset, as in
e.g.\/~\cite{energycrisis} for the case of $\mathrm{AdS}_2$, we
instead rigorously pursue a superspace approach; analyzing the
superspace geometry of the supersphere we construct in particular
the invariant vielbein and spin connection, using a
super-generalization of the left-invariant Maurer-Cartan form for
ordinary Lie groups (see Section \ref{sec:superzweibein}). Having
this information at hand we proceed to construct a
\emph{superscalar} field theory on $S^{2|2}$, which only when
written in terms of the component fields of the superscalar field
under consideration, and after integrating out the odd
coordinates, becomes a field theory on the ordinary sphere. Having
derived the component field version of the superfield action in
Section \ref{sec:fieldtheory}, we will be able to briefly discuss
supersymmetry breaking in Section \ref{sec:susybreaking}. Notably,
the superspace approach also makes it possible to derive Lagrange
equations as well as Noether's theorem for the superscalar field
itself, see Section \ref{sec:noether}.

%% file: uosp.tex
\section{The unitary orthosymplectic group}\label{sec:uosp}

\subsection{Pseudo-conjugation}
\label{sec:pseudo-conjugation} We expand an arbitrary (complex)
supernumber $z$ in terms of the generators of a Grassmann algebra
$\zeta_i$, $i=1,\ldots, N$, as
\be\label{eqn:supernum}
z=z_0+z_i\zeta_i+z_{ij}\zeta_i\zeta_j+z_{ijk}\zeta_i\zeta_j\zeta_k+\ldots
\ee
We use a subscript $0$ to denote the \emph{body} of the
supernumber, the remaining terms are called the \emph{soul}. A
supernumber is said to be \emph{even} if the above expansion does
not contain terms with an odd number of generators. The set of
even supernumbers  will be denoted by $\C_c$. A supernumber is
said to be \emph{odd} if it contains only terms with an odd number
of generators. The odd supernumbers will be denoted by $\C_a$. The
set of all supernumbers will be denoted by $\C_N$. We will
normally, however, consider the formal limit $N\to \infty$ and
denote the supernumbers by $\C_\infty$. Note also that
$\C_0\cong\C$ is precisely the set of ordinary complex numbers.

The standard extension of
ordinary complex conjugation to supernumbers is given in \cite{dewitt}. It
is defined as a map
\be
*:\left\{
\begin{array}{l}
\C_c\to\C_c \\
\C_a\to\C_a
\end{array} \right.
\ee
which agrees with complex conjugation on ordinary numbers
and satisfies the following properties
\begin{align}
(a+b)^* & = a^*+b^*, \\
(ab)^* & = b^*a^*, \\
a^{**} & = a,
\end{align}
for arbitrary supernumbers $a$ and $b$. Note that when taking the
conjugate of a product the order is reversed.
The Grassmann generators can be taken to be real with respect to this
conjugation, i.e.\/~$\zeta_i^*=\zeta_i$, and the expansion of $z^*$ is
given by
\be
z^*=z_0^*+z_i^*\zeta_i-z_{ij}^*\zeta_i\zeta_j-z_{ijk}^*\zeta_i\zeta_j\zeta_k+\ldots
\ee
Note that the minus signs are due to the reordering of the Grassmann
generators.

It is possible to define another extension of complex conjugation to
supernumbers, called \emph{pseudo-conjugation}
\cite{scheunert}. Pseudo-conjugation is defined as a map
\be\label{eqn:diamondmap}
\diamond:\left\{
\begin{array}{l}
\C_c\to\C_c \\
\C_a\to\C_a
\end{array} \right.
\ee
which agrees with complex conjugation on ordinary numbers and satisfies the following properties
\begin{align}
(a+b)^\diamond & = a^\diamond+b^\diamond, \\
(ab)^\diamond & = a^\diamond b^\diamond, \\
a^{\diamond\diamond} & = (-1)^{\epsilon_a}a \label{eqn:dbldiamond},
\end{align}
for arbitrary supernumbers $a$ and $b$, where $\epsilon_a=0$ if
$a\in\C_c$ and $\epsilon_a=1$ if $a\in\C_a$. Note that the
pseudo-conjugate does \emph{not} switch the order when applied to
a product. A consequence of this definition is that the generators
of the Grassmann algebra can no longer be described as real with
respect to pseudo-conjugation in the same way as for standard
conjugation. To see this note that if we had
$\zeta_i^\diamond=\zeta_i$ this would imply that
$\zeta_i^{\diamond\diamond}=\zeta_i$ which contradicts
\refe{eqn:dbldiamond}. In fact, a definition of how the
pseudo-conjugate acts on the Grassmann generators, which is
consistent with \refes{eqn:diamondmap}{eqn:dbldiamond}, is not
always possible. If however $N$ is even, or indeed infinite, we
can proceed as follows. Let $W$ be the $N$-dimensional vector
space of Grassmann generators. Pick a semilinear map\footnote{A
map $f$ is said to be semilinear if
$f(\boldsymbol{u}+\boldsymbol{v})=f(\boldsymbol{u})+f(\boldsymbol{v})$
and $f(\lambda \boldsymbol{v})=\lambda^*f(\boldsymbol{v})$, where
$*$ is a field automorphism, e.g.\/~complex conjugation.} $f:W\to
W$ such that $f^2=-1$, for example the matrix
\be
J=\left(\begin{array}{cc}
0 & \mathbbm{1}_{N/2} \\
-\mathbbm{1}_{N/2} & 0
\end{array}\right),
\ee
and then define $\zeta_i^\diamond=f(\zeta_i)$.
Using this definition of the pseudo-conjugate on the Grassmann
generators it is possible to write down the expansion for an arbitrary
supernumber as
\be
z^\diamond=z_0^*+z_i^*\zeta_i^\diamond+z_{ij}^*\zeta_i^\diamond\zeta_j^\diamond+z_{ijk}^*\zeta_i^\diamond\zeta_j^\diamond\zeta_k^\diamond+\ldots
\ee

%%%%%%%%%%%%%%%%%%%%%%%%%%

\subsection{Graded adjoint}\label{sec:gradedadjoint}

Using ordinary complex conjugation of supernumbers it is possible
to define an adjoint operation on \emph{pure supermatrices}. A
pure, i.e.\/~\emph{even} or \emph{odd}, $(p|q)$-dimensional
supermatrix is written in block form as
\be\label{eqn:blockform}
X  =  \left(\begin{array}{cc}
A & B \\
C & D
\end{array}\right).
\ee
The matrix is said to be even if $A\in \mbox{Mat}_{p\times
p}(\C_c)$, $B\in \mbox{Mat}_{p\times q}(\C_a)$, $C\in
\mbox{Mat}_{q\times p}(\C_a)$ and $D\in \mbox{Mat}_{q\times
q}(\C_c)$. The matrix is called odd if $A\in \mbox{Mat}_{p\times
p}(\C_a)$, $B\in \mbox{Mat}_{p\times q}(\C_c)$, $C\in
\mbox{Mat}_{q\times p}(\C_c)$ and $D\in \mbox{Mat}_{q\times
q}(\C_a)$. Here $\mbox{Mat}_{m\times n}(\mathbbm{F})$ are $m\times
n$ matrices over $\mathbbm{F}$.

The standard adjoint operation is defined, as usual, by the conjugate
transpose
\be\label{eqn:adjoint}
X^\dagger=X^{*\transpose},
\ee
or in block form
\be
\left(\begin{array}{cc}
A & B \\
C & D
\end{array}\right)^\dagger=\left(\begin{array}{cc}
A^{*\transpose} & C^{*\transpose} \\
B^{*\transpose} & D^{*\transpose}
\end{array}\right).
\ee
This satisfies the usual properties of an adjoint
\begin{align}
(XY)^\dagger & = Y^\dagger X^\dagger, \\
X^{\dagger\dagger} & = X.
\end{align}

It is also possible to use the pseudo-conjugate to construct a
\emph{graded adjoint} \cite{scheunert}. Note, however, that one
cannot construct an adjoint operation which has sensible
properties using the pseudo-conjugate combined with the ordinary
transpose, but rather one has to use the \emph{supertranspose}.
The supertranspose of a pure $(p|q)$-dimensional supermatrix is
defined by
\be
\left(\begin{array}{cc}
A & B \\
C & D
\end{array}\right)^{\stranspose}=\left(\begin{array}{cc}
A^\transpose & (-1)^{\epsilon_X}C^\transpose \\
-(-1)^{\epsilon_X}B^\transpose & D^\transpose
\end{array}\right),
\ee
where $\epsilon_X=0$ for even supermatrices, and $\epsilon_X=1$
for odd supermatrices\footnote{Note that with this definition of
the supertranspose we have that in general
$(X^{\stranspose})^{\stranspose} \ne X$, see \cite{scheunert}.}.
The graded adjoint is then defined as
\be
X^\ddagger=X^{\diamond \stranspose},
\ee
and this satisfies a graded version of the properties of the
standard adjoint
\begin{align}
(XY)^\ddagger & = (-1)^{\epsilon_X\epsilon_Y}Y^\ddagger X^\ddagger, \\
X^{\ddagger\ddagger} & = (-1)^{\epsilon_X}X.
\end{align}

We may also extend the definition of the graded adjoint to
\emph{supervectors} in a manner consistent with the definition for
supermatrices. We write a pure, i.e.\/~even or odd,
$(p|q)$-dimensional supervector as
\be
V=\left(\begin{array}{c} u \\ w
\end{array}\right).
\ee
The supervector is said to be even, i.e.\/~$\epsilon_V=0$, if
$u\in\mbox{Mat}_{p\times 1}(\C_c)$ and $w\in\mbox{Mat}_{q\times
  1}(\C_a)$. It is called odd, i.e.\/~$\epsilon_V=1$, if
$u\in\mbox{Mat}_{p\times 1}(\C_a)$ and $w\in\mbox{Mat}_{q\times
  1}(\C_c)$. We define the supertranspose of $V$ to be
\be
V^\stranspose= \left(\begin{array}{cc} u^\transpose, &
(-1)^{\epsilon_V}w^\transpose\end{array}\right)
\ee
and the graded adjoint is then defined by
\be
V^\ddagger=V^{\diamond\stranspose}.
\ee

%%%%%%%%%%%%%%%%%%%%%%%%%%

\subsection{Compact form of $O\!Sp(n|2m)$}

Using the graded adjoint one can define a compact (i.e.\/~unitary)
form of the orthosymplectic supergroup $O\!Sp(n|2m)$ which is not
possible with the ordinary adjoint.

The orthosymplectic supergroup is defined by \cite{scheunert}
\be
O\!Sp(n|2m)=\left\{ s\in PL(n|2m) : s^{\stranspose}gs=g \right\},
\ee
where $PL(n|2m)$ are the invertible even supermatrices of
dimension $(n|2m)$ and
\be
g=\left(\begin{array}{cc}
\mathbbm{1}_n & 0 \\
0 & J_{2m}
\end{array}\right),
\quad
J_{2m}=\left(\begin{array}{cc}
0 & \mathbbm{1}_m \\
-\mathbbm{1}_m & 0
\end{array}\right).
\ee
The algebra is given by
\be
osp(n|2m)=\left\{X\in pl(n|2m): X^{\stranspose}g+gX=0\right\},
\ee
where $pl(n|2m)$ is the algebra of $PL(n|2m)$.
If we write $X$ in block form, as in \refe{eqn:blockform}, then for
$X$ to be in the algebra it must satisfy
\begin{align}
A^\transpose +A & = 0, \label{eqn:algebra1}\\
B+C^\transpose J & = 0, \label{eqn:algebra2}\\
D^\transpose J+JD & = 0. \label{eqn:algebra3}
\end{align}
From \refec{eqn:algebra1}{eqn:algebra3} we see that the body of the
algebra is
\be
osp(n|2m)_0=o(n)\times sp(2m).
\ee

To find a compact form of an algebra we must first complexify it
and then impose a consistent antihermitian condition, which yields
a unitary group. For the orthosymplectic algebra the standard
adjoint of \refe{eqn:adjoint} does not give a consistent
antihermitian condition. To see this note that imposing
$X^\dagger=-X$ we find $B^{*\transpose}=-C$ and
$C^{*\transpose}=-B$. From \refe{eqn:algebra2} we have
\be
0=(B+C^\transpose J)^*J=-C^\transpose J+B
\ee
which together with \refe{eqn:algebra2} would imply $B=C=0$. This
problem is avoided by using the graded adjoint. Imposing
$X^\ddagger=-X$ we have $B^{\diamond \transpose}=C$ and
$C^{\diamond
  \transpose}=-B$. The previous argument now gives
\be
0=(B+C^\transpose J)^\diamond J=C^\transpose J+B
\ee
and hence no inconsistency.

The \emph{unitary} orthosymplectic algebra can now be defined as
\be
uosp(n|2m)=\left\{X\in osp(n|2m)\otimes\C_c: X^\ddagger=-X\right\},
\ee
and the group as
\be
U\!O\!Sp(n|2m)=\left\{s\in O\!Sp(n|2m)\otimes\C_c:
  s^\ddagger=s^{-1},\; \mbox{sdet}(s)=1 \right\},
%U\!O\!Sp(n|2m)=\left\{exp(X):X\in uosp(n|2m)\right\}.
\ee
where the \emph{superdeterminant} is defined by
\be
\label{eqn:superdet} \mbox{sdet}\left(\begin{array}{cc}
A & B \\
C & D
\end{array}\right) = \mbox{det}(A-BD^{-1}C)\mbox{det}(D)^{-1}.
\ee
Note that in the definition of $U\!O\!Sp(n|2m)$ we have imposed the condition
$\mbox{sdet}(s)=1$, hence strictly speaking we are dealing with the
\emph{special} unitary orthosymplectic group, we shall not however
refer to it as such.

%%%%%%%%%%%%%%%%%%%%%%%%%%%%%

\subsection{$U\!O\!Sp(1|2)$}

We will be interested in the particular case of $U\!O\!Sp(1|2)$. The
algebra has three even generators $J_i$, $i=0,1,2$ and two odd
generators $Q_\alpha$, $\alpha=-,+$, which can be represented as
supermatrices
\be \label{eqn:matrixrep}
J_i=\frac{i}{2}\left(\begin{array}{c|c}
0 & 0 \quad 0 \\ \hline
0 &    \\
0 & \mbox{\raisebox{1.5ex}[1.5ex]{\LARGE$\sigma_i$}}
\end{array}\right), \quad
Q_-=\frac{1}{2}\left(\begin{array}{c|cc}
0 & 0 & 1 \\ \hline
1 & 0 & 0 \\
0 & 0 & 0
\end{array}\right), \quad
Q_+=\frac{1}{2}\left(\begin{array}{c|cc}
0 & -1 & 0 \\ \hline
0 & 0 & 0 \\
1 & 0 & 0
\end{array}\right),
\ee
where $(\sigma_i)^\alpha{}_\beta$ are the standard Pauli matrices.
The generators of the algebra satisfy the following commutation
and anti-commutation relations,
\begin{align}
\lbrack J_i,J_j\rbrack & = -\epsilon_{ij}{}^k J_k, \label{eqn:jj}\\
\lbrack J_i,Q_\alpha\rbrack & =
\frac{i}{2}(\sigma_i)_{\alpha}{}^\beta Q_\beta, \label{eqn:jq}\\
\lbrack Q_\alpha, Q_\beta\rbrack & =
\frac{i}{2}(\sigma^i)_{\alpha\beta}J_i, \label{eqn:qq}
\end{align}
where $\epsilon_{ijk}$ is completely antisymmetric with
$\epsilon_{012}=1$. The indices $i,j,\ldots$ have been raised and
lowered using $\delta^{ij}=\delta_{ij}=\delta^i{}_j$, whereas
$\alpha,\beta,\ldots$ have been raised and lowered using the
antisymmetric symbols $\epsilon^{\alpha\beta}$ and
$\epsilon_{\alpha\beta}$, with $\epsilon^{-+}=\epsilon_{-+}= 1$.
The raising and lowering conventions, along with their application
to the Pauli matrices, are discussed more in Appendix
\ref{sec:epsilon}. The bracket $\lbrack\, ,\rbrack$ shall denote
the anti-commutator whenever both entries are odd, as e.g.\/~in
\refe{eqn:qq}. In any other case $\lbrack\,,\rbrack$ is to be
understood as the commutator.

The Casimir operator $C$ of $uosp(1|2)$ is given by
\begin{equation*}
C = J^i J_i - \epsilon^{\alpha \beta} Q_\alpha Q_\beta.
\end{equation*}
A general element of the algebra can be expanded as $X=\theta^i
J_i+\eta^\alpha Q_\alpha$, where $\theta^i \in\R_c$ and
$\eta^-=\eta^\diamond$, $\eta^+=\eta$, with $\eta\in\C_a$. We find
that $X$ is antihermitian as the generators satisfy the following
hermiticity properties
\begin{align}
J_i^\ddagger & = -J_i, \\
(Q^\ddagger)^\alpha & = \epsilon^{\alpha\beta}Q_\beta.
\end{align}
Note that if we naively multiplied the generators by
  a supernumber we would not obtain an antihermitian element $X$. The
  correct definition of left and right multiplication is \cite{bluebook}
\begin{align}
z\left(\begin{array}{cc} A & B \\ C & D
\end{array}\right) & = \left(\begin{array}{cc} zA & zB \\
  (-1)^{\epsilon_z}zC & (-1)^{\epsilon_z}zD \end{array}\right), \\
  \left(\begin{array}{cc} A & B \\ C & D
\end{array}\right)z & = \left(\begin{array}{cc} Az & (-1)^{\epsilon_z}Bz \\
  Cz & (-1)^{\epsilon_z}Dz \end{array}\right).
\end{align}

The general element of the group $U\!O\!Sp(1|2)$ can be represented by a
supermatrix
\renewcommand{\arraystretch}{1.2}%
\begin{align}
s(a,b,\eta) & = \left(\begin{array}{ccc}
1+\frac{1}{4}\eta\eta^\diamond  &  -\frac{1}{2}\eta  &
\frac{1}{2}\eta^\diamond \\
-\frac{1}{2}\eta^\diamond  &  1-\frac{1}{8}\eta\eta^\diamond  &  0 \\
-\frac{1}{2}\eta  &  0  &  1-\frac{1}{8}\eta\eta^\diamond
\end{array}\right)\!\!\!
\left(\begin{array}{ccc}
1 & 0 & 0 \\
0 & a & -b^\diamond \\
0 & b & a^\diamond
\end{array}\right), \label{eqn:generalelement2}
\\
& = \left(\begin{array}{ccc}
1+\frac{1}{4}\eta\eta^\diamond  &  -\frac{1}{2}(\eta a-\eta^\diamond
b)  &  \frac{1}{2}(\eta b^\diamond+\eta^\diamond a^\diamond) \\
-\frac{1}{2}\eta^\diamond  &  (1-\frac{1}{8}\eta\eta^\diamond)a  &
-(1-\frac{1}{8}\eta\eta^\diamond)b^\diamond \\
-\frac{1}{2}\eta  &  (1-\frac{1}{8}\eta\eta^\diamond)b  &
(1-\frac{1}{8}\eta\eta^\diamond)a^\diamond
\end{array}\right), \label{eqn:generalelement}
\end{align}
\renewcommand{\arraystretch}{1}%
where the parameters $a,b\in \C_c$ are constrained by
$\mbox{sdet}(s)=aa^\diamond+bb^\diamond=1$, and $\eta\in\C_a$ is
unconstrained. Note that the first matrix on the right hand side
of \refe{eqn:generalelement2} is just $\mbox{exp}(\eta^\alpha
Q_\alpha)$. The second matrix is of the form $\mbox{exp}(\theta^i
J_i)$, for some $\theta^i\in\R_c$ determining the constrained
parameters $a$ and $b$. From this we see that the body of
$uosp(1|2)$ is simply $uosp(1|2)_0=su(2)$ which will be important
in the next section.

%% file: coset.tex
\section{Constructing the supersphere}

\subsection{General coset spaces}\label{sec:generalcoset}

We shall briefly review the general formalism for
constructing spaces as coset spaces, covered in, for example, \cite{west}.

Consider a group $G$ with a subgroup $H$. We define an equivalence
relation on $G$ by
\be
g_1 \sim g_2 \iff g_2^{-1}g_1 \in H.
\ee
Each element $g\in G$ lies in an equivalence class
\be
gH\equiv\{gh:h\in H\}.
\ee
The set of all equivalence classes is the (right-)coset space $G/H$, written as
\be
G/H\equiv\{gH:g\in G\}.
\ee

We can define a projection map $\pi:G\to G/H$ by sending an
element $g\in G$ to its equivalence class $gH\in G/H$. Also, for each
point in the coset space we may choose
a particular element of $G$ which projects down to this point under
$\pi$, this group element is called a \emph{coset
  representative}.

The left action of $G$ on itself descends to an action of $G$ on
the coset space
\begin{align}
g' & : G/H \to G/H \\
& : gH \mapsto g'gH.
\end{align}
This is well defined as it is clearly independent of the coset
representative chosen.

In Section \ref{sec:superzweibein} we will introduce a vielbein and
spin connection on $G/H$ which are invariant under this left
action, and as such we will think of $G$ as the isometry group of the
coset space.

%%%%%%%%%%%%%%%%%%%%%%%%%%%%%%%%%%%%%%%%

\subsection{The sphere as a coset space}

We first review how the ordinary sphere can be constructed as the coset
space $S^2=SU(2)/U(1)$. This construction is then straightforward to
generalize to the case of the supersphere.

The group $SU(2)$ has the $2\times 2$ matrix representation
\be
s(a,b)=\left(\begin{array}{cc}
a & -b^* \\
b & a^*
\end{array}\right),
\ee
where the parameters $a,b\in\C_0$ are just ordinary complex numbers
which are constrained by $aa^*+bb^*=1$. The
matrices $s(w,0)$, with $ww^*=1$, form a $U(1)$ subgroup.
We define an equivalence relation on $SU(2)$ by multiplication on the
right with an element of this $U(1)$ subgroup.
\be
s(a,b)\sim s(a',b')=s(a,b)s(w,0).
\ee
This equivalence relation defines the coset space $SU(2)/U(1)$. The
projection map for this coset space is the standard \emph{Hopf map}, it can
be written as
\begin{align}
\pi & : SU(2)\to SU(2)/U(1) \\
& : s(a,b)\mapsto s(a,b) \hat J_0 s(a,b)^\dagger,
\end{align}
where $\hat J_0=\frac{i}{2}\sigma_0$ is the element of the algebra $su(2)$
which generates the $U(1)$ subgroup. Note we consider the
image of $\pi$ as a subset of the algebra $su(2)$, which is just
$\mathbbm{R}^3$ as a vector space. Expanding the image in coordinates
we have
\be
 s(a,b) J_0 s(a,b)^\dagger = \sum_{i=0}^2 x^i J_i,
\ee
where $x^i\in\R$. This equation leads to the constraint
\be
(x^0)^2+(x^1)^2+(x^2)^2=1,
\ee
hence the coset space $SU(2)/U(1)$ is just an ordinary sphere,
$S^2\subseteq\mathbbm{R}^3$.

%%%%%%%%%%%%%%%%%%%%%%%%%%%%

\subsection{The supersphere as a coset space}\label{sec:s22coset}

The construction of the previous section naturally generalizes to the
case of the supersphere \cite{thepaper,landimarmo}, which we will see can be
defined as the coset space $S^{2|2}=U\!O\!Sp(1|2)/\mathcal{U}(1)$.

We use the matrix representation of $U\!O\!Sp(1|2)$ defined in
\refec{eqn:generalelement2}{eqn:generalelement}. The equivalence relation on
$U\!O\!Sp(1|2)$ is given by multiplication on the right with an element of a
$\mathcal{U}(1)$ subgroup\footnote{$\mathcal{U}(1)\equiv\{w\in\C_c :
  ww^\diamond=1\}$ is the even Grassmann extension of the group
  $U(1)$.},
\be
s(a,b,\eta) \sim s(a',b',\eta')=s(a,b,\eta)s(w,0,0).
\ee
In terms of the group parameters we have,
\be a' = aw, \qquad b' = bw, \qquad \eta' =
\eta.\label{eqn:paramequiva}
\ee
This equivalence relation defines the
coset space $U\!O\!Sp(1|2)/\mathcal{U}(1)$. Note that the body of this coset
space is just $SU(2)/U(1)$, which as we showed in the previous section
is just an $S^2$. The projection map for
this coset is a supersymmetric generalization of the ordinary Hopf
map, it can be written as
\begin{align}
\pi & : U\!O\!Sp(1|2)\to U\!O\!Sp(1|2)/\mathcal{U}(1) \\
& : s(a,b,\eta)\mapsto s(a,b,\eta) J_0 s(a,b,\eta)^\ddagger.
\end{align}
Note that the image of this map is considered as a
subset of the algebra $uosp(1|2)$. Expanding the image in coordinates
we have
\be
 s(a,b,\eta) J_0 s(a,b,\eta)^\ddagger = \sum_{i=0}^2 x^i J_i +
 \sum_{\alpha=\pm} \xi^\alpha Q_\alpha ,
\ee
where $x^i\in \R_c$ and $\xi^\pm\in\C_a$. It is then possible to
solve for the coordinates in terms of the group parameters, which
yields
\begin{align}
x^0 & = (1-\frac{1}{4}\eta\eta^\diamond)(aa^\diamond-bb^\diamond) \label{eqn:x0},\\
x^1 & = (1-\frac{1}{4}\eta\eta^\diamond)(ab^\diamond+a^\diamond b), \\
x^2 & = i(1-\frac{1}{4}\eta\eta^\diamond)(ab^\diamond-a^\diamond b), \\
\xi^- & = -\frac{i}{2}(2\eta
ab^\diamond+\eta^\diamond(aa^\diamond-bb^\diamond)), \\
\xi^+ & =  - \frac{i}{2}(2\eta^\diamond a^\diamond
b-\eta(aa^\diamond-bb^\diamond)). \label{eqn:xi+}
\end{align}
These coordinates satisfy the constraint
\be\label{eqn:supersphere}
(x^0)^2+(x^1)^2+(x^2)^2-2\xi^-\xi^+=1,
\ee
which is the equation for the unit supersphere
$S^{2|2}\subseteq\mathbbm{R}^{3|2}$. Another way to think about
this equation is as a two-sphere in the even coordinates, with a
radius dependent on the odd coordinates, given by
$1+\xi^-\xi^+=1-\frac{1}{4}\eta\eta^\diamond$. It is also clear
from \refe{eqn:supersphere} that the body of the supersphere is
just an ordinary sphere, as expected.

The reality of the coordinates $x^i$ and $\xi^{\alpha}$ is defined
with respect to the pseudo-conjugate; we have $(x^i)^\diamond=x^i$
and $(\xi^-)^\diamond=-\xi^+$, ($(\xi^+)^\diamond=\xi^-$). Note
that if we expand out the coordinates in terms of the Grassmann
generators as in \refe{eqn:supernum} then these reality conditions
give the same number of constraints\footnote{Obviously for the
purposes of
  counting these constraints we must take the number of Grassmann
  generators, $N$, to be finite.} as would be obtained with standard
complex conjugation, which reduces the dimensionality down from
that of $\mathbbm{C}^{3|2}$ to that of $\mathbbm{R}^{3|2}$.

%%%%%%%%%%%%%%%%%%%%%%%%%%%

\subsection{Unconstrained coordinates}\label{sec:unconstrained}

In this section we will construct unconstrained coordinates on the
supersphere. On $S^2$ we can define, for example, polar and
stereographic coordinates and we will generalize these to
$S^{2|2}$ in the following.

We first note that the general element of $U\!O\!Sp(1|2)$ can be written as
\be\label{eqn:generalform}
s=e^{\eta^\alpha Q_\alpha}e^{-\varphi J_0}e^{-\theta J_2}e^{-\psi
J_0}.
\ee
Here $\theta,\varphi,\psi\in\R_c$, and their bodies\footnote{Here
the body of $\theta$, denoted by $\theta_0$, should not be
confused with the coordinate $\theta^0$.} are chosen to be in the
range $0\leq\theta_0\leq\pi$, $-\pi<\varphi_0\leq\pi$ and
$-\pi<\psi_0\leq\pi$. A convenient choice of coset representative
is given by taking $\psi=0$, i.e.\/~
\be\label{eqn:cosetreppolar}
L_1(\theta,\varphi,\eta,\eta^\diamond)=e^{\eta^\alpha
Q_\alpha}e^{-\varphi J_0}e^{-\theta J_2}.
\ee
Thus we have $(\theta,\varphi,\eta,\eta^\diamond)$ as coordinates
on $S^{2|2}$. The constrained coordinates
$(x^0,x^1,x^2,\xi^-,\xi^+)$ of \refes{eqn:x0}{eqn:xi+} can be
written in terms of these generalized polar coordinates as
\begin{align}
x^0 & = (1-\frac{1}{4}\eta\eta^\diamond)\cos\theta, \label{eqn:polarx0}\\
x^1 & = (1-\frac{1}{4}\eta\eta^\diamond)\sin\theta\cos\varphi, \\
x^2 & = (1-\frac{1}{4}\eta\eta^\diamond)\sin\theta\sin\varphi, \label{eqn:polarx2}\\
\xi^- & = -\frac{i}{2}(\eta
e^{-i\varphi}\sin\theta+\eta^\diamond\cos\theta), \\
\xi^+ & = -\frac{i}{2}(\eta^\diamond
e^{i\varphi}\sin\theta-\eta\cos\theta).
\end{align}
Note that the trigonometric functions for supernumbers are defined in
terms of the usual power series; the usual trigonometric identities
are satisfied if the angles are even supernumbers. Also note the
appearance of the radius factor, $1-\frac{1}{4}\eta\eta^\diamond$, in \refes{eqn:polarx0}{eqn:polarx2}.

To define a generalization of stereographic coordinates we take a
different coset representative $L_2(z,z^\diamond,\eta,\eta^\diamond)$,
which can be written in the matrix representation of
\refe{eqn:generalelement2} as
\renewcommand{\arraystretch}{1.4}%
\be
L_2 %(z,z^\diamond,\eta,\eta^\diamond)
=
\left(\begin{array}{ccc}
1+\frac{1}{4}\eta\eta^\diamond  &  -\frac{1}{2}\eta  &
\frac{1}{2}\eta^\diamond \\
-\frac{1}{2}\eta^\diamond  &  1-\frac{1}{8}\eta\eta^\diamond  &  0 \\
-\frac{1}{2}\eta  &  0  &  1-\frac{1}{8}\eta\eta^\diamond
\end{array}\right)\!\!\!\!
\left(\begin{array}{ccc}
1 & 0 & 0 \\
0 & \frac{z^\diamond}{(1+zz^\diamond)^{1/2}} & \frac{-1}{(1+zz^\diamond)^{1/2}} \\
0 & \frac{1}{(1+zz^\diamond)^{1/2}} &
\frac{z}{(1+zz^\diamond)^{1/2}}
\end{array}\right).
\ee
\renewcommand{\arraystretch}{1}%
The complex coordinate $z$ is related to the previous coordinates by
\be\label{eqn:zxphi}
z=\frac{x^1+ix^2}{1+\xi^-\xi^+-x^0} =
\frac{e^{i\varphi}\sin\theta}{1-\cos\theta},
\ee
where again the radius factor, $1+\xi^-\xi^+$, appears. We will
find later that the coordinate $\eta$ is not the most convenient
for our purposes, with hindsight we thus define a new odd
coordinate $\chi$, and its pseudo-conjugate $\chi^\diamond$, by
\be
\chi = -\frac{i}{2}(\eta^\diamond z + \eta),
\qquad
\chi^\diamond = \frac{i}{2}(\eta^\diamond - \eta z^\diamond).\label{eqn:chi}
\ee
These relations can be inverted, giving
\be
\label{eqn:etachi} \eta=\frac{2i(\chi^\diamond
z+\chi)}{(1+zz^\diamond)}, \qquad \eta^\diamond=\frac{2i(\chi
z^\diamond-\chi^\diamond)}{(1+zz^\diamond)}.
\ee
Rewriting $\eta$ in terms of $\chi$ gives us the coset
representative for the point $(z,\chi)$, which we write as
\be\label{eqn:cosetrep}
L_3(z,z^\diamond,\chi,\chi^\diamond)=L_2(z,z^\diamond,\eta(\chi,\chi^\diamond),\eta^\diamond(\chi,\chi^\diamond)).
\ee

Note that the coordinates $(z,\chi)$ cover a single
$\mathbbm{C}^{1|1}$ chart on $S^{2|2}$. From \refe{eqn:zxphi} we see
that as $\theta_0\to 0$ we have $z\to \infty$, thus these coordinates
can be viewed as generalizations of stereographic coordinates
projected from the north pole (i.e.\/~$\theta=0$).
To cover the entire supersphere we need a second coordinate patch,
which we will think of as projection from the south pole.
Away from both the north and south pole we define a new (even) complex
coordinate by $w=z^{-1}$. A coset representative for the point
$(w,\eta)$ is given by
\renewcommand{\arraystretch}{1.4}%
\be
L'_2 %(w,w^\diamond,\eta,\eta^\diamond)
=
\left(\begin{array}{ccc}
1+\frac{1}{4}\eta\eta^\diamond  &  -\frac{1}{2}\eta  &
\frac{1}{2}\eta^\diamond \\
-\frac{1}{2}\eta^\diamond  &  1-\frac{1}{8}\eta\eta^\diamond  &  0 \\
-\frac{1}{2}\eta  &  0  &  1-\frac{1}{8}\eta\eta^\diamond
\end{array}\right)\!\!\!\!
\left(\begin{array}{ccc}
1 & 0 & 0 \\
0 & \frac{1}{(1+ww^\diamond)^{1/2}} & \frac{-w}{(1+ww^\diamond)^{1/2}} \\
0 & \frac{w^\diamond}{(1+ww^\diamond)^{1/2}} &
\frac{1}{(1+ww^\diamond)^{1/2}}
\end{array}\right).
\ee
\renewcommand{\arraystretch}{1}%
This can be obtained from the coset representative $L_2$ by
multiplication on the right with
\be
s\!\left(\frac{z}{(zz^\diamond)^{1/2}},0,0\right)=\left(\begin{array}{ccc}1 & 0 & 0 \\
0 & \frac{z}{(zz^\diamond)^{1/2}} & 0 \\
0 & 0 & \frac{z^\diamond}{(zz^\diamond)^{1/2}}
\end{array}\right),
\ee
which, as it should be, is an element of the $\mathcal{U}(1)$ subgroup
of $\uosp$.
We will also need the analogue of the coordinate $\chi$ for this
patch, which we take to be
\be
\zeta=\frac{i}{2}(\eta^\diamond w+\eta).
\ee
Away from the poles, the two coordinate patches are related by the holomorphic
transformations
\be
w = \frac{1}{z}, \qquad
\zeta = -\frac{\chi}{z}.
\ee
The two patches $(z,\chi)$ and $(w,\zeta)$ taken together cover the whole
supersphere.

%%%%%%%%%%%%%%%%%%%%

\subsection{Other superspheres}
\label{sec:othersuperspheres}

At this stage we should mention that the coset space
$S^{2|2}=\uosp/\mathcal{U}(1)$ is not the only way in which a
supersphere can be defined. There are at least two other possible
coset constructions.

\begin{itemize}

\item $O\!Sp(3|2)/O\!Sp(2|2)$ --- The ordinary two-sphere can be
  constructed as the coset space $O(3)/O(2)$; since the body of
  $O\!Sp(n|2m)$ is just $O(n)\times Sp(2m)$ it is natural to consider
  the coset space $O\!Sp(3|2)/O\!Sp(2|2)$ as a supersymmetric
  generalization of this \cite{saleur}. The body of this space is clearly just the ordinary
  two-sphere.  Just as $\uosp/\mathcal{U}(1)$ is, as a subset of
  $\R^{3|2}$, given by \refe{eqn:supersphere}, so is
  $O\!Sp(3|2)/O\!Sp(2|2)$.
  Now, however, the coordinates $x^i$ and $\xi^\alpha$ are just real
  supernumbers, i.e.\/~when expanded in the Grassmann generators, as in
  \refe{eqn:supernum}, all the coefficients are real numbers.

\item $SU(2|1)/U(1|1)$ --- This construction is a generalization of
  that of the complex projective plane. The body of this coset space
  is given by $U(2)/\left(U(1)\times U(1)\right)=\mathbbm{C}P^1$. As
  the orthosymplectic groups are not used in this construction the use of
  the pseudo-conjugate  and graded adjoint is not required. This
  space, called $\mathbbm{C}P^{1|1}$, and its generalizations
  $\mathbbm{C}P^{n|m}$ are considered further in \cite{townsend}.

\end{itemize}

However, neither of these two coset spaces can naturally be
considered what one calls a \emph{superspace}. A coset space $G/H$
will be a superspace if it satisfies two conditions. Firstly, the
subgroup $H$ should be (the even Grassmann extension of) the
tangent space group of the body of the coset space. This will
correspond to a restriction of the tangent space group of a
general supermanifold. Secondly, we require that under the adjoint
action of $H$, elements of the Fermi sector\footnote{The Fermi
sector of a superalgebra is spanned by the odd generators.} of the
algebra of $G$ transform as spinors. The coset space
$\uosp/\mathcal{U}(1)$ satisfies both of these conditions: $U(1)$
is the tangent space group of the ordinary sphere, and we see from
\refe{eqn:jq} that $Q_\pm$ transform as spinors. Most other
treatments use the supersphere as a target space for some sigma
model \cite{spanishguy,saleur} and thus do not require a
superspace structure. Here we shall be treating the supersphere as
the base space for our field theories and as such require it to be
a superspace. This will be discussed more in Section
\ref{sec:superzweibein}.

%% file: isometries.tex
\section{Action of $U\!O\!Sp(1|2)$ on $S^{2|2}$}

\subsection{Transformation of the coordinates under $\uosp$}

Using the general result of Section \ref{sec:generalcoset} we see
that the left action of $U\!O\!Sp(1|2)$ is well defined on the
coset space $S^{2|2}$. First we wish to show how such a
transformation acts on the unconstrained coordinates $(z,\chi)$
which were defined in Section \ref{sec:unconstrained}. The left
action of the arbitrary element $s(c,d,\beta)\in\uosp$ transforms
the coset representative $L_3(z,z^\diamond,\chi,\chi^\diamond)$ as
\be
L_3(z,z^\diamond,\chi,\chi^\diamond)\to L_3(z',z'^\diamond,\chi',\chi'^\diamond)=s(c,d,\beta)L_3(z,z^\diamond,\chi,\chi^\diamond).
\ee
We can split the transformation as
$s(c,d,\beta)=s(1,0,\beta)s(c,d,0)$ and analyze the two parts
separately. Using \refec{eqn:generalelement2}{eqn:cosetrep} we
find, that under the action of $s(c,d,0)$ the coordinates
transform as
\begin{align}
z' & =  \frac{c^\diamond z-d}{d^\diamond z+c}, \label{eqn:zrot} \\
\chi ' & =  \frac{\chi}{d^\diamond z+c}, \label{eqn:chirot}
\end{align}
whereas under $s(1,0,\beta)$ we have
\begin{align}
z' & =  (1-\frac{i}{2}\beta^\diamond\chi)z-\frac{i}{2}\beta\chi,
\label{eqn:zsusy}\\
\chi' & =
(1+\frac{1}{8}\beta\beta^\diamond)\chi-\frac{i}{2}(\beta+\beta^\diamond
z). \label{eqn:chisusy}
\end{align}
Obviously we can take the pseudo-conjugate of these equations to find
how $z^\diamond$ and $\chi^\diamond$ transform.

Note that the group element $s(c,d,0)$ is obtained by exponentiating
just the $J_i$ generators of the $uosp(1|2)$ algebra. We also see that the form
of \refe{eqn:zrot} is that of a  M\"obius transformation corresponding
to the rotation of a sphere. We thus refer to the transformations of
\refec{eqn:zrot}{eqn:chirot} as the rotations of the supersphere.
The group element $s(1,0,\beta)$ is obtained by exponentiating only
the $Q_\alpha$ algebra generators. We will therefore refer to
\refec{eqn:zsusy}{eqn:chisusy} as the supersymmetry transformations.

For completeness we must also consider how the $(w,\zeta)$ coordinates
of the other patch transform. We find that under rotations given by
$s(c,d,0)$ we have
\begin{align}
w' & =  \frac{d^\diamond + cw}{c^\diamond-dw}, \\
\zeta' & =  \frac{\zeta}{c^\diamond-dw}.
\end{align}
Under supersymmetry transformations given by $s(1,0,\beta)$ we have
\begin{align}
w' & =  (1-\frac{i}{2}\beta\zeta)w-\frac{i}{2}\beta^\diamond\zeta, \\
\zeta' & =
(1+\frac{1}{8}\beta\beta^\diamond)\zeta+\frac{i}{2}(\beta^\diamond +
\beta w).
\end{align}
Again we may take the pseudo-conjugate of these equations to find
the transformation properties of $w^\diamond$ and $\zeta^\diamond$.

%%%%%%%%%%%%%%%%%%%%%%%%%%%%%%%%%%%%%%%%%%%

\subsection{Differential operator representation of $uosp(1|2)$}\label{sec:diffop}

We may use the transformation properties of the coordinates under
$U\!O\!Sp(1|2)$ to construct a differential operator representation of
the algebra $uosp(1|2)$.

The coordinates $(z,z^\diamond,\chi,\chi^\diamond)$ can be
represented by a single superspace coordinate $X^M$, where the
index $M=(m,\mu)$ runs over $m=z,z^\diamond$,
$\mu=\chi,\chi^\diamond$. We may then define a superscalar field
$\Phi$ on the supersphere, which is just a supernumber valued
function on $S^{2|2}$. In this coordinate patch it takes the value
$\Phi(X)$.

Now consider an infinitesimal active coordinate transformation $X\to
X+\delta X$. As discussed more in Appendix \ref{sec:superscalardensity}, we
may alternatively think of this as a transformation of the field,
$\Phi(X)$, given by
\be
\Phi(X)\to \Phi'(X)=\Phi(X-\delta X).
\ee
Expanding to first order we have
\be\label{eqn:deltaphi}
\delta\Phi(X)=-\delta X^M\partial_M\Phi(X).
\ee
For the case of an isometry we can write $\delta X^M = \delta u K^M$,
where $\delta u$ is some small parameter, and $K^M$ is a Killing
supervector. The quantity $-K^M\partial_M$ will then be the
differential operator corresponding to the isometry.

First we shall consider the rotations of
\refec{eqn:zrot}{eqn:chirot}. For a rotation generated by the
element $J_0$ we have $s(c,d,0)=e^{\theta^0J_0}$, hence
\be
c=e^{i\theta^0/2}, \qquad d=0.
\ee
Expanding \refec{eqn:zrot}{eqn:chirot} to first order in
$\theta^0$ we find
\be
\delta z=-i\theta^0z, \qquad \delta\chi=-\frac{i}{2}\theta^0\chi,
\ee
$\delta z^\diamond$ and $\delta \chi^\diamond$ are obtained by
taking the pseudo-conjugate of these equations. Substituting into
\refe{eqn:deltaphi} gives us the differential operator
corresponding to $J_0$, namely
\be\label{eqn:J0}
\tilde J_0=i\left[z\frac{\partial}{\partial
    z} - z^\diamond\frac{\partial}{\partial
    z^\diamond} + \frac{1}{2}\chi\frac{\partial}{\partial\chi} -
  \frac{1}{2}\chi^\diamond\frac{\partial}{\partial\chi^\diamond}
  \right].
\ee
A similar argument leads to the differential operators for $J_1$ and
$J_2$,
\begin{align}
\tilde J_1 & =  \frac{i}{2}\left[(1-z^2)\frac{\partial}{\partial
    z} - (1-{z^\diamond}^2)\frac{\partial}{\partial
    z^\diamond} - z\chi\frac{\partial}{\partial\chi} +
  z^\diamond\chi^\diamond\frac{\partial}{\partial\chi^\diamond}
  \right], \\
\tilde J_2 & =  -\frac{1}{2}\left[(1+z^2)\frac{\partial}{\partial
    z} + (1+{z^\diamond}^2)\frac{\partial}{\partial
    z^\diamond} + z\chi\frac{\partial}{\partial\chi} +
  z^\diamond\chi^\diamond\frac{\partial}{\partial\chi^\diamond}
  \right].
\end{align}

Now consider the supersymmetry transformations of
\refec{eqn:zsusy}{eqn:chisusy}. Expanding these to first order in
$\beta$ and $\beta^\diamond$, and substituting into
\refe{eqn:deltaphi} we find the differential operators
corresponding to $Q_-$ and $Q_+$,
\begin{align}
\tilde Q_- & =  \frac{i}{2}\left[\chi z\frac{\partial}{\partial z} -
\chi^\diamond
  \frac{\partial}{\partial z^\diamond} +
  z\frac{\partial}{\partial\chi}-\frac{\partial}{\partial\chi^\diamond} \right], \\
\tilde Q_+ & =  \frac{i}{2}\left[\chi\frac{\partial}{\partial z} +
  \chi^\diamond z^\diamond
  \frac{\partial}{\partial z^\diamond} + \frac{\partial}{\partial\chi}
  + z^\diamond\frac{\partial}{\partial\chi^\diamond} \right]. \label{eqn:Q+}
\end{align}
It is straightforward to verify that the generators of
\refes{eqn:J0}{eqn:Q+} satisfy the $uosp(1|2)$ algebra. As stated
earlier they are of the form $-K_p^M\partial_M$, where
$p=0,1,2,-,+$ labels the generators. This allows us to read off
the Killing supervectors $K_p^M$ of the supersphere.

In order to construct a superfield theory on $S^{2|2}$ we first
have to introduce the invariant vielbein and spin connection,
which we do next.

%% file: superzweibein.tex
\section{Coset space geometry}
\label{sec:superzweibein}
\subsection{Vielbein and spin connection for reductive coset spaces}
Consider some Lie group $G$, a subgroup $H$ of $G$ and the space
of right cosets $G/H = \{gH: g \in G\}$. The Lie algebra
$\mathfrak{h}$ of $H$ is spanned by the generators $H_I \in
\mathfrak{h}$, $I = 1, \ldots, \mbox{dim}H$. Let the remaining
generators of the Lie algebra $\mathfrak{g}$ of $G$ span
 $\mathfrak{k} \subseteq \mathfrak{g}$. We shall
denote these remaining generators by $K_A \in \mathfrak{k}$, $A = 1,
\ldots, \mbox{dim}G - \mbox{dim}H$. As a vector space we then have
\begin{equation}
\mathfrak{g} = \mathfrak{h} \oplus \mathfrak{k}.
\end{equation}

The structure constants of $G$ are defined by
\begin{align}
[H_I, H_J] & =  f_{IJ}{}^K H_K \\
\left[H_I, K_A\right] & =  f_{IA}{}^J H_J + f_{IA}{}^B K_B \\
\left[K_A, K_B\right] & =  f_{AB}{}^J H_J + f_{AB}{}^C K_C.
\end{align}
If $\mathfrak{k}$ can be chosen such that the structure constants
$f_{IA}{}^J$ vanish, the coset space $G/H$ is said to be
\emph{reductive}.

Suppose now that the coset manifold $G/H$ is parameterized by coordinates
$Y^M$, $M=1,\ldots , \mbox{dim}G - \mbox{dim}H$, and so the coset
representative may be written $L(Y^M)$. For reductive coset spaces we can
then define an invariant vielbein $E^A$ and spin connection $\omega^I$ by
\begin{equation}
\label{eqn:maurer-cartan} L^{-1}(Y) dL(Y) = E^A K_A + \omega^I H_I
\end{equation}
which is a generalization of the left-invariant Maurer--Cartan form for
Lie groups. Here $L(Y)$ is assumed to be in a matrix representation.

Note that these are indeed \emph{invariant} one-forms since under
a left action of $g \in G$ on the coset space we have
\begin{align}
L & \mapsto  g L \\
L^{-1}dL & \mapsto  ( L^{-1}g^{-1})\,d(gL) = L^{-1}dL,
\end{align}
where $g$ is constant on the coset space. Hence we can think of
this action as an isometry.

In contrast, under a right action of $h^{-1} \in H$ on the coset
space we find
\begin{align}
\label{eqn:rightaction}
L & \mapsto  Lh^{-1} \\
L^{-1}dL & \mapsto  hL^{-1}d(Lh^{-1}) = h(L^{-1}dL)h^{-1} + hdh^{-1}
\end{align}
and hence
\begin{equation}
E^AK_A + \omega^I H_I \mapsto  \underbrace{h(E^A K_A)h^{-1}}_{\in \:
\mathfrak{k}} + \underbrace{h(\omega^I H_I)h^{-1} + hdh^{-1}}_{\in \:
\mathfrak{h}}.
\end{equation}
Here $h = h(Y)$, i.e.\/~$h$ is not necessarily constant on the
coset space, but is rather a local transformation. Note that
$h(E^A K_A)h^{-1} \in \mathfrak{k}$ is only true for reductive
coset spaces. Thus we have
\begin{alignat}{2}
E^A K_A & \mapsto  E\,'{}^A K_A \; &  = &  \; E^A (h K_A h^{-1}) \label{eqn:gaugetrafo1} \\
\label{eqn:gaugetrafo2}\omega^I H_I & \mapsto  \omega\,'{}^I H_I &
= &  \; \omega^I (h H_I h^{-1}) + hdh^{-1}.
\end{alignat}
We can rewrite this using the co-adjoint representation\footnote{Obviously
this can also be written using the adjoint representation, see
e.g.\/~\cite{coset}.} of $G$, i.e.\/~$g \mapsto R_p{}^{q}(g)$, which is
defined as
\begin{equation}
\label{eqn:coadjointrep} g^{-1}T_{p} g = R_p{}^{q}(g)\,T_q,
\end{equation}
where $T_{p}$, $p = 1,\ldots,\mbox{dim}G$, are the generators of
$\mathfrak{g}$. Thus we have
\begin{equation}
h K_A h^{-1} = R_A{}^B\!(h^{-1})\,K_B
\end{equation}
and so we can alternatively write
\begin{equation}
\label{eqn:gaugetrafosuperzweibein} E^A \mapsto E\,'{}^A = E^B
R_B{}^A(h^{-1}).
\end{equation}
Rewriting Eqn.\/~(\ref{eqn:gaugetrafo2}) in the co-adjoint
representation we find
\begin{equation}
\omega'{}^I (\tilde{H}_I)_A{}^B = \omega^I R_A{}^C(h)
(\tilde{H}_I)_C{}^D R_D{}^B(h^{-1}) + R_A{}^C(h)dR_C{}^B(h^{-1}),
\end{equation}
where $\tilde{H}_I$ denotes the generator $H_I$ in the co-adjoint
representation. Defining $\Omega_A{}^B = \omega{}^I
(\tilde{H}_I)_A{}^B$ we can finally write
Eqn.\/~(\ref{eqn:gaugetrafo2}) as
\begin{equation}
\Omega_A{}^B \mapsto \Omega'_A{}^B = R_A{}^C(h) \Omega_C{}^D
R_D{}^B(h^{-1}) + R_A{}^C(h)dR_C{}^B(h^{-1}).
\end{equation}
In this sense the right action of $h^{-1}$ on the coset space,
defined in Eqn.\/~(\ref{eqn:rightaction}), can be regarded as a
local gauge transformation acting on the tangent space.

%%%%%%%%%%%%%%%%%%%%%%%%%%%%%%%%%%%%%%%%%%%%%%%%%%%%%%%%%%%%%

\subsection{Vielbein and spin connection for $S^{2|2}$}
\label{sec:vielbeinspin}

We will now derive the superzweibein and spin connection for
$S^{2|2}$ following the construction given in the previous
section.

As mentioned in Section \ref{sec:s22coset} the supersphere
$S^{2|2}$ is, as a coset space, given by $S^{2|2} = U\!O\!Sp(1|2)
/\mathcal{U}(1)$. As before we will split up the generators of $G
= U\!O\!Sp(1|2)$ into the generator of the subgroup $H =
\mathcal{U}(1)$, which we take to be $J_0$, and the remaining
generators $K_A$, $A = (a, \alpha)$, which are given by $J_a,
Q_\alpha$, with $a = 1,2$, $\alpha = -,+$, see
\refes{eqn:jj}{eqn:qq}. In this case we have
--- apart from $S^{2|2}$ being a reductive coset
space\footnote{Note that $[H, K] \subseteq K$} --- the additional
structure that
\begin{align}
[H,Q] & \subseteq  Q \\
{[H, J]} & \subseteq  J
\end{align}
hence
\begin{alignat}{2}
h^{-1} Q_\alpha h  = &  \hspace{1ex} R_{\alpha}{}^A(h) K_A \, & = & \; R_\alpha{}^\beta(h) Q_\beta\\
h^{-1} J_a h   = & \hspace{1ex} R_a{}^A(h) K_A \, & = & \;
R_a{}^b(h) J_b.
\end{alignat}
Thus $R_A{}^B(h)$ takes block diagonal form
\begin{equation}
R_A{}^B(h) = \left(%
\begin{array}{c|c}
  R_a{}^b(h) & 0 \\\hline
 \mbox{\raisebox{-0.5ex}[-0.5ex]{0}} & \mbox{\raisebox{-0.5ex}[-0.5ex]{$R_\alpha{}^\beta(h)$}} \\
\end{array}%
\right).
\end{equation}

Using the matrix representation of the $U\!O\!Sp(1|2)$ algebra, see
\refe{eqn:matrixrep}, we find for $R_a{}^b(h)$ and $R_\alpha{}^\beta(h)$,
respectively
\begin{align}
R_a{}^b(h) & =  \left(%
\begin{array}{cc}
\cos \varphi & \sin \varphi\\
- \sin \varphi & \cos \varphi
\end{array}%
\right)\\
R_\alpha{}^\beta(h) & =  \left(%
\begin{array}{cc}
e^{-i\varphi/2} & 0 \\
0 & e^{i \varphi/2}
\end{array}%
\right).
\end{align}
We see that tangent supervectors $V^A$ belong to a (completely) reducible
representation of the tangent space group; the components $V^a$ transform
in the vector representation, whereas the components $V^\alpha$ transform
in the corresponding spinor representation of $\mathcal{U}(1)$. In this
sense we are dealing with a superspace rather than just a supermanifold
(see Section \ref{sec:othersuperspheres}).

To construct the superzweibein and spin connection in the particular case
of $\uosp/\mathcal{U}(1)$ we have to choose an appropriate coset
representative. This is given by
\begin{equation}
L_1(\theta,\phi,\eta,\eta^\diamond)=e^{\eta^\alpha
Q_\alpha}e^{-\phi J_0}e^{-\theta J_2},
\end{equation}
as defined in \refe{eqn:cosetreppolar}. In matrix form (see
\refe{eqn:generalelement2}) we have
\renewcommand{\arraystretch}{1.15}%
\begin{equation}
L_1(\theta,\phi,\eta,\eta^\diamond)  = \left(\begin{array}{ccc}
1+\frac{1}{4}\eta\eta^\diamond & -\frac{1}{2}\eta  &
\frac{1}{2}\eta^\diamond \\
-\frac{1}{2}\eta^\diamond  &  1-\frac{1}{8}\eta\eta^\diamond  &  0 \\
-\frac{1}{2}\eta  &  0  &  1-\frac{1}{8}\eta\eta^\diamond
\end{array}\right)\!\!\!
\left(\begin{array}{ccc} 1 & 0 & 0 \\
0 & a & -b^\diamond \\
0 & b & a^\diamond
\end{array}\right),
\end{equation}
\renewcommand{\arraystretch}{1}%
where here $a(\theta, \varphi) = e^{-i\varphi/2} \cos{\frac{\theta}{2}}$
and $b(\theta, \varphi) = e^{i \varphi /2} \sin{\frac{\theta}{2}}$.
According to the general formalism derived in the previous section, the
superzweibein and spin connection for $S^{2|2}$ as the coset space can be
derived from the generalized Maurer-Cartan one-form,
\refe{eqn:maurer-cartan},
\begin{equation}
L_1{}^{-1}(\theta,\phi,\eta,\eta^\diamond)
dL_1(\theta,\phi,\eta,\eta^\diamond) = E^A K_A + \omega^I H_I,
\end{equation}
with $H_I = J_0$ and $K_A = (J_a, Q_\alpha)$, $a = 1,2$, $\alpha =
-, +$. This way we obtain the superzweibein and spin connection in
(super)-polar coordinates. Their explicit form is given in
Appendix \ref{sec:appendixpolar}.

Using instead the coset representative defined in
\refe{eqn:cosetrep} we find for the superzweibein in complex
(stereographic) coordinates\footnote{Note that the two coset
representatives, (\refec{eqn:cosetreppolar}{eqn:cosetrep}), differ
by a gauge transformation only. Thus, the superzweibein in complex
coordinates can be derived from the one in polar coordinates by
means of a gauge transformation, see
\refe{eqn:gaugetrafosuperzweibein}.}
{\renewcommand{\arraystretch}{1.9}%
\begin{equation}
(E_M{}^A) =
\left(%
\! \begin{array}{cc|cc}
  \frac{-i}{1 + z z^\diamond + \chi \chi^\diamond} & \frac{1}{1 + z z^\diamond + \chi \chi^\diamond} &
  \frac{- 2i (\chi z^\diamond - \chi^\diamond)}{(1 + z z^\diamond)^{3/2}} & 0 \\
  \frac{i}{1 + z z^\diamond + \chi \chi^\diamond} &  \frac{1}{1 + z z^\diamond + \chi \chi^\diamond} & 0 &
  \frac{- 2i (\chi^\diamond z + \chi)}{(1 + z z^\diamond)^{3/2}}
  \\ \hline
  \frac{-i \chi}{1 + z z^\diamond} &  \frac{\chi}{1 + z z^\diamond} & \frac{2i}{(1 + z z^\diamond - \chi \chi^\diamond)^{1/2}}  &  0 \\
  \frac{i \chi^\diamond}{1 + z z^\diamond} & \frac{\chi^\diamond}{1 + z z^\diamond} & 0 & \frac{2i}{(1 + z z^\diamond - \chi \chi^\diamond)^{1/2}}  \\
\end{array}%
\!\right),
\end{equation}}%
where the index $M$, as before, runs over $z, z^\diamond, \chi,
\chi^\diamond$. For the inverse super\-zweibein, which we will
make extensive use of later, we have
\renewcommand{\arraystretch}{2}%
{\setlength{\arraycolsep}{0.5mm}%
\begin{equation}
(E_A{}^M) = {\mbox{\scriptsize
$\left(%
\! \begin{array}{cc|cc}
  \frac{i}{2}(1 + z z^\diamond) & - \frac{i}{2}(1 + z z^\diamond) &
  \frac{i}{2} (\chi  z^\diamond - \chi^\diamond) & - \frac{i}{2} (\chi^\diamond z  +  \chi)\\
  \frac{1}{2}(1+  z z^\diamond) & \frac{1}{2}(1 +z z^\diamond) & \frac{1}{2} (\chi z^\diamond - \chi^\diamond)&
  \frac{1}{2} (\chi^\diamond z + \chi)
  \\\hline
  \frac{i}{2}(1 + z z^\diamond)^{1/2}\chi &  0 & - \frac{i}{2}(1 + z z^\diamond + \chi \chi^\diamond)^{1/2}  &  0 \\
  0  & \frac{i}{2}(1 + z z^\diamond)^{1/2}\chi^\diamond & 0 & -\frac{i}{2}(1 + z z^\diamond+ \chi \chi^\diamond)^{1/2}  \\
\end{array}%
\! \right)$}}.
\end{equation}}\renewcommand{\arraystretch}{1}%
In order to construct a superfield Lagrangian later on we will
make especial use of $E_-$ and $E_+$, which we can read off from
$(E_A{}^M)$ above. We have
\begin{align}
E_- & = \frac{i}{2}(1 + z z^\diamond + \chi
\chi^\diamond)^{\frac{1}{2}}(\chi
\partial_z - \partial_\chi) \\
E_+ & = \frac{i}{2}(1 + z z^\diamond + \chi
\chi^\diamond)^{\frac{1}{2}}(\chi^\diamond
\partial_{z^\diamond} - \partial_{\chi^\diamond}).
\end{align}
The superdeterminant, (cf. \refe{eqn:superdet}), of $(E_M{}^A)$ is
given by
\begin{equation}
\label{eqn:sdet} E \equiv \sdet(E_M{}^A) = \frac{i}{2}\;
\frac{1}{1 + z z^\diamond + \chi \chi^\diamond} = \frac{i}{2}\;
\frac{1 + z z^\diamond - \chi \chi^\diamond}{(1 + z
z^\diamond)^{2}}.
\end{equation}
Finally, we have for the spin connection in complex coordinates
\begin{align}
\label{eqn:spinconnectionzchi} \omega^0 & = \frac{i}{1 + z
z^\diamond + \chi \chi^\diamond}(z^\diamond dz - z dz^\diamond +
d\chi \chi^\diamond + d\chi^\diamond \chi)\\ \nonumber {} & = -
\frac{1}{2}(z^\diamond + z) E^1 + \frac{i}{2}(z^\diamond - z) E^2
\\
& \hspace{2.5ex} + \frac{1}{2} \frac{\chi z^\diamond - \chi^\diamond}{(1 +
z z^\diamond)^{1/2}} E^- - \frac{1}{2} \frac{\chi^\diamond z + \chi}{(1 +
z z^\diamond)^{1/2}} E^+
\end{align}
hence in the co-adjoint representation
\begin{equation}
\Omega_B{}^C = \omega^0 (J_0)_B{}^C,
\end{equation}
where
\renewcommand{\arraystretch}{1.2}%
\begin{equation}
(J_0)_B{}^C =
\left(%
\begin{array}{cc|cc}
  0 & 1 & 0 & 0 \\
  -1 & 0 & 0 & 0 \\ \hline
  0 & 0 & -\frac{i}{2} & 0 \\
  0 & 0 & 0 & \frac{i}{2} \\
\end{array}%
\right).
\end{equation}
\renewcommand{\arraystretch}{1.2}%
Note that the body of $\Omega_\alpha{}^\beta$ is given by
\begin{equation}
\label{eqn:bodyspinconnection} \Omega_\alpha{}^\beta {\Big|}_0=
\frac{i}{1 + z_0 z_0^*}(z_0^* dz_0 - z_0 dz_0^*)
(J_0)_\alpha{}^\beta,
\end{equation}
which matches the result expected for the ordinary sphere. Similar
expressions for the superzweibein, its dual and the spin
connection can be obtained for the $(w, \zeta)$ coordinate patch
(see Section \ref{sec:unconstrained}). They are given in Appendix
\ref{sec:appendixwzeta}.

The results developed in this section can be used to define a
covariant derivative on the supersphere. This will be given by
\begin{align}
\mathcal{D}_A & = E_A{}^M (\partial_M + \omega^0_M J_0) = E_A +
\Omega_A,
\end{align}
with $\Omega_A = E_A{}^M\omega^0_M J_0$ and where $J_0$ is taken
to be in the representation appropriate to the field being acted
on.

%%%%%%%%%%%%%%%%%%%%%%%%%%%%%%%%%%%%%%%%%%%%%%%
\subsection{Torsion and curvature of $S^{2|2}$}

We are now in the position to calculate the torsion components for
the supersphere and hence --- by Dragon's theorem \cite{dragon}
--- the curvature components. This can be done using the fact that
the (anti-)commutator of two covariant derivatives is determined
in terms of the supertorsion $T_{AB}{}^C$ and the supercurvature
$R_{AB}$ as follows
\begin{equation}
{[\mathcal{D}_A, \mathcal{D}_B]}
 = T_{AB}{}^C \mathcal{D}_C +
R_{AB}.
\end{equation}
Here, both the torsion and the curvature are two-forms which have
the following symmetry properties
\begin{align}
T_{AB}{}^C & = - (-1)^{\epsilon_A \epsilon_B}T_{BA}{}^C,\\
R_{AB} & = - (-1)^{\epsilon_A \epsilon_B} R_{BA},
\end{align}
with
\begin{equation}
\epsilon_A =
\left\{%
\begin{array}{ll}
    0 \hspace{2ex} \mbox{if} \; A=a \\
    1  \hspace{2ex} \mbox{if} \; A=\alpha  \\
\end{array}%
\right. .
\end{equation}
It is convenient to directly express the torsion and curvature
components in terms of the superzweibein and spin connection.
Defining the so-called \emph{anholonomy} coefficients
$\mathcal{C}_{AB}{}^C$ by
\begin{equation}
{[E_A,E_B]}
 = \mathcal{C}_{AB}{}^C E_C,
\end{equation}
we have
\begin{align}
T_{AB}{}^C & = \mathcal{C}_{AB}{}^C + \Omega_{AB}{}^C -
(-1)^{\epsilon_A \epsilon_B} \Omega_{BA}{}^C \\
R_{ABC}{}^D & = E_A \Omega_{BC}{}^D + \Omega_{AC}{}^E
\Omega_{BE}{}^D - (-1)^{\epsilon_A \epsilon_B}(A \leftrightarrow
B) - \mathcal{C}_{AB}{}^E \Omega_{EC}{}^D.
\end{align}
Note that as a result of the Bianchi identities and of the
restricted choice of tangent space group the curvature is
completely determined in terms of the torsion. This is known as
\emph{Dragon's theorem}.

The only non-vanishing torsion components are given by
\begin{align}
T_{\alpha \beta}{}^a & = \frac{i}{2} (\sigma^a)_{\alpha{\beta}} \\
T_{\alpha a}{}^\beta & = - \frac{i}{2}
(\sigma_a)_\alpha{}^{\beta},
\end{align}
where the invariant tensor $(\sigma^a)^{\alpha}{}_{\beta}$ is
given in Appendix \ref{sec:epsilon}. Note that even for flat
superspace one finds non-zero torsion components $T_{\alpha
\beta}{}^a$. Since the curvature is completely determined in terms
of the torsion we must therefore expect some other non-vanishing
torsion components in the case of $S^{2|2}$, which is a curved
superspace. Thus it is not surprising that we encounter the
additional torsion components $T_{\alpha a}{}^\beta$.

For the only non-vanishing curvature components we find
\begin{alignat}{2}
R_{12B}{}^C & = - R_{21B}{}^C & = &\hspace{1ex} (J_0)_B{}^C \\
R_{-+B}{}^C & = R_{+-B}{}^C & = &\hspace{1ex}-\frac{i}{2}(J_0)_B{}^C.
\end{alignat}
Note that the only non-zero components of the body of the
curvature tensor, $R_{abc}{}^d$, are given by $R_{12a}{}^b = -
R_{12a}{}^b = (J_0)_a{}^b$, which matches the result for the
ordinary sphere.

In the following we will use the geometric structure developed in
this section to formulate scalar field theories on $S^{2|2}$.
Before we do so, however, we will discuss superscalar fields on
the supersphere and their transformation properties under
isometries.

%% file: superfields.tex
\section{Superfields on the supersphere}\label{sec:superfields}

\subsection{Component fields}

In Section \ref{sec:diffop} we defined a superscalar field,
$\Phi$, on the supersphere. Working in the $(z,\chi)$ coordinate
patch we can perform an expansion in the $\chi$ and
$\chi^\diamond$ variables, giving
\be
\Phi=A(z,z^\diamond)+\chi\psi_\chi(z,z^\diamond)+
\chi^\diamond\psi_{\chi^\diamond}(z,z^\diamond)+\chi\chi^\diamond
F(z,z^\diamond).
\ee
 The fields $A$, $\psi_\chi$, $\psi_{\chi^\diamond}$ and $F$ are
 called the component fields of $\Phi$, and are functions of $z$ and
 $z^\diamond$ only. $F$ is often referred to as the
 \emph{auxiliary} field.

Since we know how the superfield $\Phi$ transforms under
isometries (see Eqn. (\ref{eqn:deltaphi})), it is possible to
derive how the component fields transform. For example, under the
action of $J_0$ we have $\delta\Phi=\theta^0\tilde J_0\Phi$, which
gives
\begin{align}
\delta A & =  i\theta^0 \left(z\partial_z -
z^\diamond\partial_{z^\diamond}\right)A, \label{eqn:componentrot1} \\
\delta \psi_\chi & =  i\theta^0 \left(z\partial_z -
z^\diamond\partial_{z^\diamond}+\frac{1}{2}\right)\psi_\chi, \label{eqn:componentrot2} \\
\delta \psi_{\chi^\diamond} & =  i\theta^0 \left(z\partial_z -
z^\diamond\partial_{z^\diamond}-\frac{1}{2}\right)\psi_{\chi^\diamond}, \label{eqn:componentrot3} \\
\delta F & =  i\theta^0 \left(z\partial_z -
z^\diamond\partial_{z^\diamond}\right)F.\label{eqn:componentrot4}
\end{align}
Similar expressions for the transformation properties under $J_1$
and $J_2$ can also be found. An identical argument gives the
transformation of the component fields under the supersymmetry
transformation $\delta\Phi=\beta^\alpha\tilde Q_\alpha\Phi$. We
find
\begin{align}
\delta A & =  \frac{i}{2}\left((\beta^\diamond z+\beta)\psi_\chi + (\beta
z^\diamond-\beta^\diamond)\psi_{\chi^\diamond} \right),
\label{eqn:componentsusy1} \\
\delta \psi_\chi & =  \frac{i}{2}\left((\beta z^\diamond-\beta^\diamond)F- (\beta^\diamond z+\beta)\partial_zA\right), \label{eqn:componentsusy2} \\
\delta \psi_{\chi^\diamond} & =  -\frac{i}{2}\left((\beta^\diamond
z+\beta)F +(\beta z^\diamond-\beta^\diamond)\partial_{z^\diamond}A\right),\label{eqn:componentsusy3}  \\
\delta F & =  \frac{i}{2}\left( (\beta^\diamond
z+\beta)\partial_z\psi_{\chi^\diamond}-(\beta
z^\diamond-\beta^\diamond)\partial_{z^\diamond}\psi_\chi \right).
\label{eqn:componentsusy4}
\end{align}

It is possible to put these equations in a more familiar form
by rewriting them using \emph{Killing spinors}, which we do next.

%%%%%%%%%%%%%%%%%%%%%%%%%55

\subsection{Killing spinors}\label{sec:killing}
In order to define Killing spinors we must first introduce some
more notation concerning the geometry of $S^{2|0}$, the even
Grassmann extension of the ordinary two-sphere. The gamma
matrices, $\gamma^m$, $m=z,z^\diamond$, for $S^{2|0}$, can be
taken to be
\begin{align}
\gamma^z & =  -i(1+zz^\diamond)\left(\begin{array}{cc} 0 & 1 \\ 0 & 0
\end{array}\right), \\
\gamma^{z^\diamond} & =  i(1+zz^\diamond)\left(\begin{array}{cc} 0 &
  0 \\ 1 & 0 \end{array}\right).
\end{align}
These satisfy $\{\gamma^m,\gamma^n\}=2g^{mn}$ where the metric
$g_{mn}$ has the following non-zero components
\be
g_{zz^\diamond}=g_{z^\diamond z}=\frac{2}{(1+zz^\diamond)^2}.
\ee
As we can see from \refe{eqn:spinconnectionzchi}, the restriction
of the spin connection $\omega^0$ from the supersphere to
$S^{2|0}$ is given by
\begin{equation}
\omega \equiv \omega^0 \Big|_{\chi, \chi^\diamond = 0} = \omega_z
dz + \omega_{z^\diamond} dz^\diamond
 = \frac{i}{(1+zz^\diamond)}(z^\diamond
  dz-zdz^\diamond)\left(\begin{array}{cc} -\frac{i}{2} & 0 \\ 0 &
  \frac{i}{2} \end{array}\right).
\end{equation}
This allows us to define the covariant derivative
  $D_m=\partial_m+\omega_m$.

Killing spinors on $S^{2|0}$ are defined by (see
\cite{killingspinors})
\be
\label{eqn:killingspinor}
D_m\boldsymbol{\epsilon}=\frac{i}{2}\kappa\gamma_m\boldsymbol{\epsilon},
\ee
where $\kappa=\pm 1$.
A solution to this equation with $\kappa=-1$ reads
\be\label{eqn:killingspinorsolution}
\boldsymbol{\epsilon}=\frac{1}{2(1+zz^\diamond)^\frac{1}{2}}\left(\begin{array}{c}
  \beta^\diamond - \beta z^\diamond \\
  \beta^\diamond z+\beta
\end{array}\right),
\ee
where $\beta\in\C_a$ is some arbitrary constant.

In order to rewrite \refes{eqn:componentsusy1}{eqn:componentsusy4}
using Killing spinors we also need to introduce a new set of
component fields, which are obtained from the superfield $\Phi$.
In the case of the spinor and auxiliary fields this will require
the use of the covariant derivative. We define
\begin{align}
\tilde A &= \Phi\Big|_{\chi, \chi^\diamond = 0}, \\
\label{eqn:psialpha}\psi_\alpha &= 2 \left(\mathcal{D}_\alpha
\Phi\right)\Big|_{\chi,
\chi^\diamond = 0}, \\
F_{\alpha\beta} &= -  \left((\mathcal{D}_\alpha \mathcal{D}_\beta
- \mathcal{D}_\beta \mathcal{D}_\alpha)\Phi \right) \Big|_{\chi,
\chi^\diamond = 0}.
\end{align}
We can use $F_{\alpha\beta}$ to alternatively define
\begin{equation}
\tilde{F} = \epsilon^{\alpha\beta}F_{\alpha\beta},
\end{equation}
where $\epsilon^{-+} = 1$.

The set of fields given by $\tilde{A}$, $\psi_-$, $\psi_+$ and
$\tilde{F}$ turns out to be a conformal rescaling of the original
component fields defined in the previous section. We find
\begin{align}
\tilde A & =  A, \label{eqn:conformalA}\\
\psi_- & =  -i(1+zz^\diamond)^\frac{1}{2}\psi_\chi, \label{eqn:conformalPsi-}\\
\psi_+ & = -i(1+zz^\diamond)^\frac{1}{2}\psi_{\chi^\diamond}, \label{eqn:conformalPsi+}\\
\tilde F & =  -(1+zz^\diamond)F\label{eqn:conformalF}.
\end{align}

Note that from \refe{eqn:psialpha} we see immediately that the
fields $\psi_-$ and $\psi_+$, carrying the tangent space index
$\alpha$, indeed transform as spinors under the action of the
tangent space group $\mathcal{U}(1)$. The components $\psi_-$ and
$\psi_+$ can be grouped into a two-component spinor,
$\boldsymbol{\psi}$, as
\be
\boldsymbol{\psi}=\left(\begin{array}{c} \psi_- \\ \psi_+
\end{array}\right).
\ee

Using these results we can rewrite the transformation of the
component fields under the supersymmetry transformations, given in
\refes{eqn:componentsusy1}{eqn:componentsusy4}, in the more
compact form
\begin{align}
\delta \tilde A & =  \boldsymbol{\epsilon}^\ddagger\boldsymbol{\psi} \\
\delta \boldsymbol{\psi} & =  (-i\partial\!\!\!/ \tilde A +\tilde
F)\boldsymbol{\epsilon} \\
\delta \tilde F & =  -i\boldsymbol{\epsilon}^\ddagger
D\!\!\!\!/\boldsymbol{\psi},
\end{align}
where the spinors $\boldsymbol{\epsilon}$ and $\boldsymbol{\psi}$
are considered as $(0|2)$-dimensional even supervectors in order
to define their graded adjoints (see Section
\ref{sec:gradedadjoint}). These equations should be compared with
standard results, for instance in \cite{energycrisis}. Note that
here the graded adjoint plays the role of the Dirac conjugate.

%% file: fieldtheory.tex
\section{Scalar field actions on $S^{2|2}$}
\label{sec:fieldtheory}
\subsection{Kinetic part of superfield action}
We are now in the position to write down a Lagrangian in terms of
some superscalar field $\Phi$. Remember that we can expand
$\Phi(z, \chi)$ in terms of the $\chi$ variables as
\begin{equation*}
\Phi=A(z,z^\diamond)+\chi\psi_\chi(z,z^\diamond)+\chi^\diamond\psi_{\chi^\diamond}(z,z^\diamond)+\chi\chi^\diamond
F(z,z^\diamond).
\end{equation*}
Here we want to restrict our attention to (pseudo-)real
superfields only. We therefore impose the reality condition
\begin{equation}
\Phi^\diamond = \Phi,
\end{equation}
which reads in terms of the component fields
\begin{align}
A^\diamond & =  A \\
\left(\psi_{\chi^\diamond}\right)^\diamond & =  - \psi_\chi \\
\left( \psi_\chi\right)^\diamond & =  \psi_{\chi^\diamond} \\
F^\diamond & =  F.
\end{align}

Let us consider the following kinetic
Lagrangian\footnote{Obviously here $L$ should not be confused with
the coset representative used earlier.}${}^,$\footnote{Since we
are dealing with a Euclidean field theory, this would perhaps be
more accurately denoted as the \emph{gradient} term of the
Lagrangian.} for the superscalar field $\Phi$
\begin{equation}
\label{eqn:lagrangian} L = \mathcal{D}_- \Phi \; \mathcal{D}_+
\Phi = E_- \Phi \; E_+ \Phi .
\end{equation}
In order to write down an action on the supersphere we will need
the invariant volume form $dz dz^\diamond d\chi d\chi^\diamond E$,
with $E = \sdet(E_M{}^A)$ as in \refe{eqn:sdet}. We thus have for
the action
\begin{equation}
\label{eqn:kineticterm} I_\mathrm{kin}
 = \int dz dz^\diamond d\chi d\chi^\diamond E \; E_- \Phi \; E_+ \Phi
\end{equation}
This will be invariant under supersymmetry transformations, as
long as the Lagrangian $L$ transforms as a scalar, e.g.\/~as
$\Phi$. This is the case, provided that under a supersymmetry
transformation with parameter $\beta$, we have
\begin{equation}
L \rightarrow e^{\beta^\alpha \tilde Q_\alpha} L.
\end{equation}
To check this, note that under a supersymmetry transformation with
small $\beta$ we have
\begin{equation}
\delta L = E_- \Phi E_+ (\beta^\alpha \tilde Q_\alpha \Phi) + E_-
(\beta^\alpha \tilde Q_\alpha \Phi) E_+ \Phi.
\end{equation}
Now using the fact that
\begin{alignat*}{2}
{[E_-, \tilde Q_-]} & = - \frac{i}{4} \chi E_-
& \qquad {[E_+, \tilde Q_-]} & = \frac{i}{4} \chi E_+ \\
{[E_-, \tilde Q_+]} & = \frac{i}{4} \chi^\diamond E_- & \qquad
{[E_+, \tilde Q_+]} & = -\frac{i}{4} \chi^\diamond E_+
\end{alignat*}
we find that $L$ indeed transforms as a scalar under supersymmetry
transformations
\begin{equation}
\delta L = \beta^\alpha \tilde Q_\alpha L.
\end{equation}
Similarly the action will be invariant under rotations if the
Lagrangian transforms as
\be
L \rightarrow e^{\theta^i \tilde J_i} L.
\ee
Under rotations, for small $\theta^i$, we have
\begin{equation}
\delta L = E_- \Phi E_+ (\theta^i \tilde J_i \Phi) + E_- (\theta^i
\tilde J_i \Phi) E_+ \Phi,
\end{equation}
which we can rewrite using
\begin{alignat*}{2}
{[E_-, \tilde J_0]} & = \frac{i}{2} E_- & \quad {[E_+, \tilde J_0]} & = -\frac{i}{2} E_+\\
{[E_-, \tilde J_1]} & = -\frac{i}{4}(z + z^\diamond) E_- & \quad
{[E_+, \tilde J_1]} & =
\frac{i}{4}(z +  z^\diamond) E_+ \\
{[E_-, \tilde J_2]} & = -\frac{1}{4}(z - z^\diamond) E_- & \quad
{[E_+, \tilde J_2]} & = \frac{1}{4}(z - z^\diamond) E_+ .
\end{alignat*}
Doing so we find
\be
\delta L = \theta^i \tilde J_i L.
\ee
Thus the action is invariant not only under supersymmetry
transformations but also under rotations.

Let us rewrite the kinetic part of the superfield action in terms of
component fields. To do so first note that we can write the Lagrangian as
\begin{equation}
\begin{split}
L(\Phi) & = E_- \Phi E_+ \Phi \\
& = - \frac{1}{4} (1 + z z^\diamond + \chi \chi^\diamond) \left[
(\chi \partial_z - \partial_\chi) \Phi \right]\left[
(\chi^\diamond
\partial_{z^\diamond}  - \partial_{\chi^\diamond}) \Phi \right],
\end{split}
\end{equation}
and thus we have for the Lagrangian density $\mathcal{L}(\Phi)$
\begin{equation}
\begin{split}
\label{eqn:lagrangiandensity}
\mathcal{L}(\Phi) & = E L(\Phi)\\
& = - \frac{i}{8} \left[ (\chi
\partial_z - \partial_\chi) \Phi \right]\left[ (\chi^\diamond
\partial_{z^\diamond}  - \partial_{\chi^\diamond}) \Phi \right].
\end{split}
\end{equation}

Expanding $\mathcal{L}(\Phi)$ in terms of the $\chi$ variables we
need to keep track only of terms proportional to $\chi
\chi^\diamond$, as these are the only ones which will survive the
Grassmann integration over $\chi$ and $\chi^\diamond$ in the
action. We have
\begin{equation}
\mathcal{L}(\Phi){\Big|}_{\chi \chi^\diamond} = - \frac{i}{8}\left(
\partial_zA \partial_{z^\diamond}A + \psi_\chi \partial_{z^\diamond} \psi_\chi +
\psi_{\chi^\diamond} \partial_{z} \psi_{\chi^\diamond} +
F^2\right).
\end{equation}
Hence we find for the action in terms of the component fields
after integrating out the $\chi$, $\chi^\diamond$ dependence
\begin{equation}
I_{\mathrm{kin}} =  \frac{i}{8} \int dz dz^\diamond \left(
\partial_zA \partial_{z^\diamond}A + \psi_\chi \partial_{z^\diamond} \psi_\chi +
\psi_{\chi^\diamond} \partial_{z} \psi_{\chi^\diamond} +
F^2\right).
\end{equation}
Note that had we used the $\eta$ coordinates instead of the $\chi$
coordinates the action would not have taken this simple form. Note
further that the kinetic part of the component field action is
conformally invariant, see Appendix \ref{sec:conformal}.

For the Euler-Lagrange equations we find
\begin{align}
\partial_z \partial_{z^\diamond} A & = 0 \\
\label{eqn:fieldeqpsichi}
\partial_{z^\diamond} \psi_\chi & = 0\\
\partial_z \psi_{\chi^\diamond} & = 0 \\
F & = 0.
\end{align}
These imply that $A$ is a harmonic function of $z$ and
$z^\diamond$, $\psi_\chi$ is a holomorphic function and
$\psi_{\chi^\diamond}$ is an anti-holomorphic function of $z$.
Thus if we insist on boundedness of the solutions, $A$ as well as
$\psi_\chi$ and $\psi_{\chi^\diamond}$ are constant in this
coordinate patch. Remember, however, that only the two coordinate
patches $(z, \chi)$ and $(w, \zeta)$ taken together cover the
whole sphere, see Section \ref{sec:unconstrained}. Thus, in order
to make a global statement, we also have to consider the field
equations following from the action written in the $(w, \zeta)$
patch. To do so, first note that we can rewrite the superfield
$\Phi$ in terms of the $w$ and $\zeta$ coordinates as
\begin{align*}
\Phi(z, \chi) & = A(z) + \chi \psi_\chi(z) + \chi^\diamond
\psi_{\chi^\diamond}(z) + \chi \chi^\diamond F(z)\\
& = A(z) - \frac{\zeta}{w} \psi_\chi(z) -
\frac{\zeta^\diamond}{w^\diamond} \psi_{\chi^\diamond}(z) + \frac{\zeta
\zeta^\diamond}{ww^\diamond} F(z).
\end{align*}
Then defining the fields
\begin{align}
\label{eqn:Ahat}
\hat{A}(w) & = A(z) \\
\label{eqn:psizetahat}
\psi_\zeta(w) & = - z\psi_\chi(z) \\
\label{eqn:psizetadiamondhat}
\psi_{\zeta^\diamond}(w) & = -z^\diamond \psi_{\chi^\diamond}(z)\\
\label{eqn:Fhat} \hat{F}(w) & = z z^\diamond F(z),
\end{align}
we have
\begin{equation}
\Phi(w, \zeta) \equiv \hat{A}(w) + \zeta \psi_\zeta(w) + \zeta^\diamond
\psi_{\zeta^\diamond}(w) + \zeta \zeta^\diamond \hat{F}(w).
\end{equation}
Using the inverse superzweibein in the $(w, \zeta)$ coordinate patch, see
\refe{eqn:inversesuper2beinwzeta}, we find for the Lagrangian
(\refe{eqn:lagrangian})
\begin{equation}
E_-\Phi E_+ \Phi = E'_-\Phi E'_+ \Phi
\end{equation}
and hence for the action in terms of the component fields defined in
\refes{eqn:Ahat}{eqn:Fhat}
\begin{equation}
I_{\mathrm{kin}} =  \frac{i}{8} \int dw dw^\diamond \left(
\partial_w\hat{A} \partial_{w^\diamond}\hat{A} + \psi_\zeta \partial_{w^\diamond} \psi_\zeta +
\psi_{\zeta^\diamond} \partial_{w} \psi_{\zeta^\diamond} +
\hat{F}^2\right).
\end{equation}
The Euler-Lagrange equations following from this action are
\begin{align}
\partial_w \partial_{w^\diamond}\hat{A} &  = 0\\
\partial_w \psi_{\zeta^\diamond} & = 0\\
\label{eqn:fieldeqpsizeta}
\partial_{w^\diamond} \psi_\zeta & = 0 \\
\hat{F} & = 0.
\end{align}
Now \refe{eqn:fieldeqpsizeta}, for example, implies that
$\psi_\zeta$ is a holomorphic function of $w$. If, however, we
insist also on boundedness of the solution we have  --- since
$\psi_\zeta = - z \psi_\chi(z)$ and since \refe{eqn:fieldeqpsichi}
implies that $\psi_\chi$ is constant --- that both $\psi_\chi(z)$
and $\psi_\zeta(w)$ must be zero. An analogous argument shows that
also $\psi_{\chi^\diamond}(z)$ and $\psi_{\zeta^\diamond}(w)$ must
be taken to be zero.
%%%%%%%%%%%%%%%%%%%%%%%%%%%%%%%%%%%%%%%%%%%%%%%%%%%%%%%%%%%%%%%%%%%%%%%%%%%%%%%%%%%%%%%

\subsection{Full superfield action}

Now let us add a potential term to the kinetic part of the
superfield action given in \refe{eqn:kineticterm}. This will allow
us later to study supersymmetry breaking in this theory. Note that
adding a potential term breaks the conformal invariance of the
action.

The potential part of the superfield action will be taken to be
\begin{equation}
\label{eqn:potentialterm} I_{\mathrm{pot}} = \frac{1}{4} \int dz
dz^\diamond d\chi d\chi^\diamond E \; U(\Phi),
\end{equation}
with $U(\Phi)$ some super-potential. When expanding $U(\Phi)$ in
terms of the odd variables one should note that, since
\begin{equation*}
E = \frac{i}{2}\frac{1 + z z^\diamond - \chi \chi^\diamond}{(1 + z
z^\diamond)^{2}},
\end{equation*}
the only terms contributing to the action after integrating out the
$\chi$, $\chi^\diamond$ dependence will be the ones proportional to 1 and
$\chi \chi^\diamond$. Keeping this in mind we write
\begin{align}
\nonumber U(\Phi) & = U(A + \chi \psi_\chi + \chi^\diamond
\psi_{\chi^\diamond} +
\chi\chi^\diamond F) \nonumber \\
    & = U(A) +\chi\chi^\diamond \left(F U'(A)  - \psi_\chi \psi_{\chi^\diamond} U''(A)
  \right) + \ldots,
\end{align}
where the dots stand for the terms proportional to $\chi$ and
$\chi^\diamond$, respectively. Thus we can rewrite $I_{\mathrm{pot}}$ in
terms of the component fields as
\begin{equation}
I_{\mathrm{pot}} = \frac{i}{8} \int dz dz^\diamond \left( \frac{U(A)}{(1 +
zz^\diamond)^2} - \frac{F U'(A) - \psi_\chi \psi_{\chi^\diamond} U''(A)}{1
+ z z^\diamond} \right).
\end{equation}

The full action $I = I_{\mathrm{kin}} + I_{\mathrm{pot}}$ in terms of the
component fields is then given by
\begin{equation}
\begin{split}
I  & = \frac{i}{8} \int dz dz^\diamond \bigg(\partial_zA
\partial_{z^\diamond}A + \psi_\chi \partial_{z^\diamond} \psi_\chi +
\psi_{\chi^\diamond} \partial_{z} \psi_{\chi^\diamond} + F^2 \\
\label{eqn:fullaction} &  \hspace{18ex}  + \frac{U(A)}{(1 +
zz^\diamond)^2} - \frac{F U'(A) - \psi_\chi \psi_{\chi^\diamond} U''(A)}{1
+ z z^\diamond}  \bigg).
\end{split}
\end{equation}
The Euler-Lagrange equations corresponding to the full action can
be found in Appendix \ref{sec:appendixeqofmotion}. Note that we
can check the invariance of the action under rotations and
supersymmetry transformations explicitly using the transformation
laws given in \refes{eqn:componentrot1}{eqn:componentsusy4}.

Seeing as $F$ is just an auxiliary field we may eliminate it from
the action. The field equation for $F$ is purely algebraic, we
have
\begin{equation*}
F = \frac{1}{2} \frac{U'(A)}{1 + z z^\diamond},
\end{equation*}
and thus eliminating it from the action we find
\begin{equation}
\begin{split}
I & = \frac{i}{8} \int dz dz^\diamond \bigg(\partial_zA
\partial_{z^\diamond}A + \psi_\chi \partial_{z^\diamond} \psi_\chi +
\psi_{\chi^\diamond} \partial_{z} \psi_{\chi^\diamond} \\
&  \hspace{13ex}  + \frac{U(A) - \frac{1}{4}\left(U'(A)\right)^2}{(1 + z
z^\diamond)^2} + \frac{U''(A)}{1 + z z^\diamond} \psi_\chi
\psi_{\chi^\diamond}\bigg).
\end{split}
\end{equation}
For later convenience we define the effective potential $V$
by\footnote{Note that the factor $(1 + zz^\diamond)^{-2}$
contributes to the invariant volume element of $S^{2|0}$ and as
such is not part of the effective potential.}
\begin{equation}
\label{eqn:effectivepotential} V(A) = U(A) -
\frac{1}{4}\left(U'(A)\right)^2.
\end{equation}
Note that the effective potential will be unbounded from below
whenever $U(A)$ is given by a polynomial of degree greater than
two. However, there exist non-polynomial choices of the potential
$U(A)$, for example a Gaussian, which lead to effective potentials
that are bounded from below.

The truncated supersymmetry transformations are
\begin{align}
\delta A & = \frac{i}{2} \left((\beta^\diamond z+\beta)\psi_\chi + (\beta
z^\diamond-\beta^\diamond)\psi_{\chi^\diamond} \right)
\\
\delta \psi_\chi & =  \frac{i}{2}\left((\beta z^\diamond-\beta^\diamond)\frac{1}{2} \frac{U'(A)}{1 + z z^\diamond}
- (\beta^\diamond z+\beta)\partial_zA\right) \\
\delta \psi_{\chi^\diamond} & =  -\frac{i}{2}\left((\beta^\diamond
z+\beta)\frac{1}{2} \frac{U'(A)}{1 + z z^\diamond} + (\beta
z^\diamond-\beta^\diamond)\partial_{z^\diamond}A\right).
\end{align}
Note that the truncated action will be invariant under these supersymmetry
transformations. However, the truncated transformations will not close
unless we impose the field equations, i.e.\/~the commutator of two
supersymmetry transformations will give a rotation only on-shell.

%% file: susybreaking.tex
\section{Supersymmetry breaking}
\label{sec:susybreaking} In this section we will investigate
supersymmetry breaking in this model for different choices of the
potential $U(\Phi)$. In order to do so let us consider an $SO(3)$
invariant classical vacuum solution given by $A = \mbox{constant}$
and $\psi_\chi = \psi_{\chi^\diamond} = 0$. Under supersymmetry
this solution transforms as
\begin{align}
\delta A & = 0\\
\delta \psi_\chi &  = \frac{i}{4}(\beta z^\diamond - \beta^\diamond)
\frac{U'(A)}{1 + zz^\diamond}\\
\delta \psi_{\chi^\diamond} &  = - \frac{i}{4}(\beta^\diamond z +
\beta) \frac{U'(A)}{1 + zz^\diamond}
\end{align}
Thus this solution will be supersymmetry preserving if $U'(A) = 0$,
i.e.\/~if $F = 0$. On the other hand $F \ne 0$ indicates states of broken
supersymmetry.

Note that vacuum solutions correspond to critical points of the effective
potential $V$, given in \refe{eqn:effectivepotential}. Since
\begin{align}
V'(A) & = U'(A)\left(1 - \frac{1}{2}U''(A) \right)
\end{align}
we have two types of stationary points, namely $U'(A)  = 0$ and $U''(A) =
2$, the former corresponding to states with unbroken supersymmetry, the
latter corresponding to states for which supersymmetry is possibly broken.

As a first example consider the potential $U(A) = mA^2$, where $m
\in \mathbbm{R}_0$ is some constant parameter\footnote{Here
$\mathbbm{R}_0 \cong\mathbbm{R}$ denotes the body of
$\mathbbm{R}_c$.}. We shall look for critical points of the
effective potential, which is
\begin{equation} V(A) = (m - m^2)A^2.
\end{equation}
Note that if $m = 0$ or 1 the effective potential is identically
zero. In the case of $m > 1$ or $m < 0$ there exists neither a
global nor a local minimum. If, however, $0 < m < 1$ the potential
possesses a global minimum at $A = 0$. As this implies that $U'(A)
= 2mA = 0$, supersymmetry will be preserved for this solution.

As a second example we will consider the potential
\begin{equation}
U(A) = \frac{1}{3} g A^3 + \lambda A,
\end{equation}
with $g, \lambda \in \mathbbm{R}_0$ constant. The extrema of the effective
potential
\begin{equation}
\label{eqn:quarticeffectivepotential} V(A) = \lambda A +
\frac{1}{3} g A^3 - \frac{1}{4}(g A^2 + \lambda)^2
\end{equation}
are given by
\begin{alignat}{2}
U'(A)  = g A^2 + \lambda & = 0 \quad & \Rightarrow \quad A & = \pm \sqrt{\frac{-\lambda}{g}}\\
U''(A)  = 2gA & = 2 \quad & \Rightarrow \quad A & = \frac{1}{g}.
\end{alignat}
In order to decide whether we can have stable supersymmetry
preserving vacuum solutions, we need to know for which parameter
values $A = \pm \sqrt{-\lambda/g}$ correspond to local minima.
Thus we need to investigate $V''(A)$ at these points. We have
\begin{align}
V''(A) & = U''(A)\left(1 - \frac{1}{2}U''(A) \right) - \frac{1}{2}
U'(A)U'''(A) \\
& = - 2 \sqrt{- \lambda g}\left(\sqrt{- \lambda g} \mp 1\right).
\end{align}
One has to distinguish between four different
cases.
\begin{figure}[t]
\begin{center}
\psfrag{1}{(\textbf{a})} \psfrag{2}{(\textbf{b})}
\psfrag{3}{(\textbf{c})} \psfrag{4}{(\textbf{d})}
\psfrag{-A1}{\scriptsize{$- A_1$}} \psfrag{A1}{\scriptsize{$A_1$}}
\psfrag{-A2}{ \scriptsize{$- A_2$}}
\psfrag{A2}{\scriptsize{$A_2$}} \psfrag{V}{\scriptsize{$V(A)$}}
\psfrag{A}{\scriptsize{$A$}}\epsfig{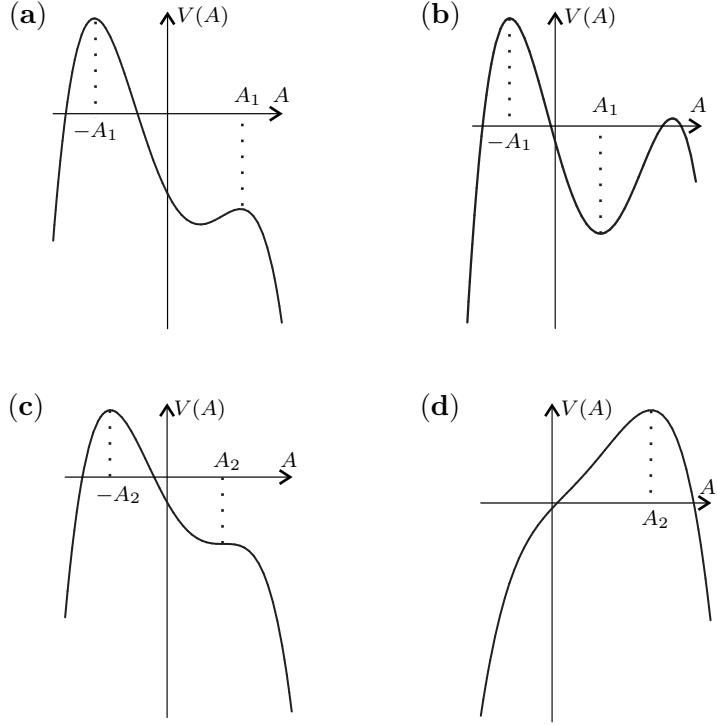}
\end{center}
\caption[]{A sketch of the effective potential $V(A) = \lambda A +
\frac{1}{3} g A^3 - \frac{1}{4}(g A^2 + \lambda)^2$ for the four
different cases discussed in the text. Here $A_1 = \sqrt{-
\lambda/g}$ and $A_2 = 1/g$.} \label{fig:superpotentials}
\end{figure}
\begin{itemize}
\item Suppose $\sqrt{- \lambda g} > 1$. In this case $V''(A) < 0$ for both
the roots $A = \pm \sqrt{-\lambda/g}$, hence $U''(A) = 2$ must correspond
to the local minimum. Thus for this vacuum solution supersymmetry will be
broken (see Fig.\/~\ref{fig:superpotentials}a).
\item Suppose $0 < \sqrt{- \lambda g} < 1$. In this case one of the roots
$A = \pm \sqrt{-\lambda/g}$ will correspond to a local maximum the other
to a local minimum. Thus there exists a supersymmetry preserving vacuum
solution (see Fig.\/~\ref{fig:superpotentials}b).
\item Suppose $- \lambda g = 1$. Then $A = \pm \sqrt{-\lambda/g} = \pm
1/g$ implies that one of the two roots corresponds to $V''(A) = 0$, the
other to a maximum. Thus there exists no stable supersymmetry preserving
vacuum state (see Fig.\/~\ref{fig:superpotentials}c).
\item Suppose $\lambda g > 0$. There is no solution to $U'(A) = 0$, hence
supersymmetry will be broken. However, in this case $V(A)$ has a single
maximum at $A = 1/g$ and thus any vacuum solution will be unstable anyway
(see Fig.\/~\ref{fig:superpotentials}d).
\end{itemize}
Note, however, that the effective potential of
\refe{eqn:quarticeffectivepotential} is unbounded from below and
as such exhibits only local minima. Therefore there do not exist
true vacuum solutions.

%% file: noether.tex
\section{Conserved currents from superfield formalism}
\label{sec:noether} Using the superfield formalism we will derive
in this section a supersymmetric generalization of the
energy-momentum tensor.

In order to do so consider some superfield Lagrangian density
$\mathcal{L} = \mathcal{L}(\Phi, \partial \Phi)$. Remember that a
coordinate transformation $X^M \rightarrow X^M + \delta X^M$ is
realized on superscalar fields as (see \refe{eqn:deltaphi}) $\Phi
\rightarrow \Phi(X - \delta X) = \Phi(X) + \delta \Phi(X)$ where
$\delta \Phi = - \delta X^M
\partial_M \Phi$. Note that in the case of an isometry, as we shall assume here,
we have $\delta X^M = \delta u K^M$, with $K^M$ a Killing
supervector.

Similarly we find that the Lagrangian density transforms under an
isometry as
\begin{equation}
\delta \mathcal{L} = - (-1)^M\partial_M \left(\delta X^M
\mathcal{L}\right).
\end{equation}
For a derivation of this result see Appendix \ref{sec:superscalardensity}.
On the other hand we find that the change in the Lagrangian density
obtained by varying the fields is\footnote{Note that the superzweibein is
invariant under an isometry, thus the variation of $\mathcal{L}$ with
respect to the superzweibein is zero.}
\begin{align}
\delta \mathcal{L} & = \delta \Phi \frac{\partial \mathcal{L}}{\partial
\Phi} + \left(\partial_M \delta \Phi \right) \frac{\partial
\mathcal{L}}{\partial (\partial_M \Phi)} \nonumber\\
\label{eqn:changeinLbyfields} & = \delta \Phi \left(\frac{\partial
\mathcal{L}}{\partial \Phi} -
\partial_M \left(\frac{\partial \mathcal{L}}{\partial (\partial_M \Phi)} \right)
\right) + \partial_M \left(\delta \Phi \frac{\partial
\mathcal{L}}{\partial (\partial_M \Phi)} \right).
\end{align}
From this we see that the Euler-Lagrange equations are
\begin{equation}
\label{eqn:fieldeqsuperfields} \frac{\partial \mathcal{L}}{\partial \Phi}
-
\partial_M \left(\frac{\partial \mathcal{L}}{\partial (\partial_M \Phi)} \right) =
0.
\end{equation}
Thus if we impose the field equations the first term in
\refe{eqn:changeinLbyfields} vanishes and we can set the remaining
term equal to $- (-1)^M\partial_M \left(\delta X^M
\mathcal{L}\right)$. Then using $\delta \Phi = - \delta
X^M\partial_M \Phi$ we find
\begin{equation}
\label{eqn:currentconservation}
\partial_M \left(\delta X^N \left((-1)^M  \delta_N{}^M \mathcal{L} - \partial_N
\Phi \frac{\partial \mathcal{L}}{\partial (\partial_M \Phi)}
\right)\right)  =  0.
\end{equation}
We are now in the position to define the super energy-momentum
tensor $\mathcal{T}_N{}^M$
\begin{equation}
\mathcal{T}_N{}^M = (-1)^M  \delta_N{}^M \mathcal{L} - \partial_N \Phi
\frac{\partial \mathcal{L}}{\partial (\partial_M \Phi)}.
\end{equation}
The corresponding super Noether current $\mathcal{J}^M$ is then defined by
\begin{equation}
\mathcal{J}^M = K^N \mathcal{T}_N{}^M.
\end{equation}
By means of \refe{eqn:currentconservation} $\mathcal{J}^M$ will
satisfy the super conservation law
\begin{equation}
\label{eqn:conservation}
\partial_M \mathcal{J}^M = 0.
\end{equation}

Let us now consider the specific Lagrangian density for $S^{2|2}$
given by (see \refec{eqn:kineticterm}{eqn:potentialterm})
\begin{equation}
\mathcal{L} = E \left(E_- \Phi E_+ \Phi + \frac{1}{4}U(\Phi) \right).
\end{equation}
One can check that the field equations given by
\refe{eqn:fieldeqsuperfields} indeed coincide --- when written in
terms of the component fields --- with the field equations given
in Appendix \ref{sec:appendixeqofmotion}, which were directly
derived from the action in terms of the component fields.

For the super energy-momentum tensor we find in this case
\begin{equation}
\begin{split}
\mathcal{T}_N{}^M & = (-1)^M  \delta_N{}^M  E \left( E_-\Phi E_+ \Phi +
\frac{1}{4}U(\Phi) \right)\\
& \hspace{3ex} - \partial_N \Phi E \left(E_-{}^M E_+ \Phi + (-1)^M E_-
\Phi E_+{}^M \right).
\end{split}
\end{equation}
The supercurrents are given by
\begin{equation}
\mathcal{J}_p{}^M = K_p{}^N \mathcal{T}_N{}^M,
\end{equation}
with $p = 0,1,2,-,+$, as before. The Killing supervectors
$K_p{}^N$ can be read off from \refes{eqn:J0}{eqn:Q+}. Note that
by taking the $\chi \chi^\diamond$ component of the conservation
equation, \refe{eqn:conservation}, we find a conservation equation
purely in $z$
\begin{equation}
\label{eqn:conservationeqcomponents} \left(\partial_M
\mathcal{J}_p{}^M\right) {\Big|}_{\chi \chi^\diamond} =
(\partial_z \mathcal{J}_p{}^z +
\partial_{z^\diamond} \mathcal{J}_p{}^{z^\diamond}){\Big|}_{\chi
\chi^\diamond} = (\partial_m \mathcal{J}_p{}^m){\Big|}_{\chi
\chi^\diamond} = 0,
\end{equation}
as both $\partial_\chi \mathcal{J}_p{}^\chi$ and
$\partial_{\chi^\diamond} \mathcal{J}_p{}^{\chi^\diamond}$ do not
contribute a $\chi \chi^\diamond$ term. It will turn out that it
is this $\chi \chi^\diamond$ contribution to the conservation
equation that gives rise to the familiar energy-momentum tensor
and fermionic currents, which can alternatively be derived
directly from the action in terms of the component fields.
Considering other components of the conservation equation, say the
$\chi$ component, we find
\begin{align}
(\partial_z \mathcal{J}_p{}^z +
\partial_{z^\diamond} \mathcal{J}_p{}^{z^\diamond}){\Big|}_{\chi} & =
- (\partial_{\chi^\diamond}
\mathcal{J}_p{}^{\chi^\diamond}){\Big|}_{\chi}.
\end{align}
Note that this also is a conservation equation purely in $z$.
However, the term on the right hand side of the equation, $-
(\partial_{\chi^\diamond}
\mathcal{J}_p{}^{\chi^\diamond}){\big|}_{\chi}$, which does not
involve any derivatives with respect to $z$, must be understood as
some kind of source term. Yet, the interpretation of these
additional conservation equations remains unclear.

Now let us consider the currents $\mathcal{J}_i{}^m$, $i = 0,1,2$,
in more detail. We have
\begin{equation*}
\mathcal{J}_i{}^m = K_i{}^N \mathcal{T}_N{}^m = K_i{}^n
\mathcal{T}_n{}^m + K_i{}^{\mu} \mathcal{T}_{\mu}{}^m.
\end{equation*}
By direct calculation one finds that the components
$\mathcal{T}_{\chi}{}^m$ are proportional to $\chi$ and similarly
the components $\mathcal{T}_{\chi^\diamond}{}^m$ are proportional
to $\chi^\diamond$. Now, as also $K_i{}^\chi$ is proportional to
$\chi$ and similarly $K_i{}^{\chi^\diamond}$ is proportional to
$\chi^\diamond$ the above equation simplifies to
\begin{equation}
\mathcal{J}_i{}^m = K_i{}^n \mathcal{T}_n{}^m.
\end{equation}
Note that $K_i{}^n = K_i{}^n {\big|}_{\chi \chi^\diamond} \equiv
k_i{}^n$ correspond to the usual Killing vectors on the sphere
$S^{2|0}$
\begin{align}
(k_0^m) & = i\left(z, -z^\diamond \right),\\
(k_1^m) & = \frac{i}{2} \left((1 - z^2),-(1 -
{z^\diamond}^2) \right),\\
(k_2^m) & = -\frac{1}{2} \left((1 + z^2), (1 + {z^\diamond}^2)
\right).
\end{align}
Defining $j_i{}^m$ and $t_m{}^n$ as the $\chi \chi^\diamond$
components of $\mathcal{J}_i{}^m$ and $\mathcal{T}_m{}^n$,
respectively
\begin{equation*}
j_i{}^m \equiv \mathcal{J}_i{}^m{\Big|}_{\chi \chi^\diamond},
\qquad t_m{}^n \equiv \mathcal{T}_m{}^n{\Big|}_{\chi
\chi^\diamond}
\end{equation*}
we can rewrite the conservation equation,
\refe{eqn:conservationeqcomponents}, for the bosonic currents
$\mathcal{J}_i{}^m$ as
\begin{equation}
\partial_m j_i{}^m = \partial_m (k_i{}^n t_n{}^m) = 0.
\end{equation}
For $t_m{}^n$ we find in terms of the conformally rescaled fields
$\tilde{A}$, $\psi_-$, $\psi_+$, as given in
\refes{eqn:conformalA}{eqn:conformalPsi+},
\begin{equation}
\begin{split}
t_{mn} & = \frac{i}{8} \sqrt{|g|} \bigg(\partial_m
\tilde{A}\partial_n\tilde{A} + \frac{i}{2} \psi^\ddag \gamma_m
\partial_n \psi \\
& \hspace{10ex} - g_{mn}\left[\frac{1}{2}(\partial \tilde{A})^2 -
\frac{1}{8}U'(\tilde{A})^2 + \frac{1}{2}U(\tilde{A})\right]\bigg),
\end{split}
\end{equation}
where the index has been lowered using the metric $g_{mn}$. Note
that the auxiliary field $\tilde{F}$ has been eliminated.

We shall now consider the $\chi\chi^\diamond$ contribution to the
currents $\mathcal{J}_\alpha{}^m$, $\alpha = -,+$. Let us define
\begin{equation}
j_\alpha{}^m \equiv \mathcal{J}_\alpha{}^m \Big|_{\chi
\chi^\diamond}
\end{equation}
and also
\begin{align}
j^m \equiv \beta^\diamond j_-{}^m + \beta j_+{}^m.
\end{align}
From \refe{eqn:conservationeqcomponents} we see that $j^m$
satisfies the conservation equation
\begin{equation}
\partial_m j^m = 0.
\end{equation}
Rewriting this fermionic current in terms of the rescaled fields
$\tilde{A}$, $\psi_-$, $\psi_+$, as we did before in the case of
$t_{nm}$, we find
\begin{equation}
j^m = \frac{i}{8} \sqrt{|g|} \boldsymbol{\epsilon}^\ddagger \left(
\partial\!\!\!/ \tilde A + \frac{i}{2} U'(\tilde{A})\right) \gamma^m
\boldsymbol{\psi},
\end{equation}
where $\boldsymbol{\epsilon}$ is the Killing spinor defined in
\refe{eqn:killingspinorsolution}.

%% file: conclusion.tex
\section{Conclusions and Outlook}
We have shown how to construct the supersphere $S^{2|2}$ as the
coset space $\uosp/\mathcal{U}(1)$, analogous to the construction
of flat superspace as the super Poincar\'e group quotiented by the
Lorentz group. The definition of $\uosp$, which is the isometry
group of the supersphere, required the notions of
pseudo-conjugation and graded adjoint.

The coset space $\uosp/\mathcal{U}(1)$ has the structure of a
superspace, rather than just being a supermanifold as is the case
for other coset space definitions of the supersphere. This allowed
us to consider the supersphere as a base space for a superscalar
field theory. As $S^{2|2}$ is an example of a curved superspace on
which we have \emph{rigid} supersymmetry transformations,
i.e.\/~the supersymmetry parameter is not position dependent, the
theory we constructed exhibits global supersymmetry. Upon
integrating out the odd coordinate dependence, this superscalar
field theory becomes a supersymmetric theory on the ordinary
sphere with a scalar, spinor and auxiliary field. Choosing a
polynomial potential we saw that solutions at local minima may
break supersymmetry, provided certain conditions are met. Also
recall that, contrary to what is expected, the effective potential
for this model is not typically bounded from below. This appears
to be due to the Euclidean nature of the theory. However, as we
pointed out, non-polynomial potentials can be found which will
exhibit global minima and thus true vacuum solutions.

Using the superfield formalism we were able to derive
Euler-Lagrange equations and Noether's theorem for the superscalar
field $\Phi$ itself, starting from some general superfield
Lagrangian density $\mathcal{L}(\Phi,\partial \Phi)$. When
applying Euler-Lagrange equations to the specific Lagrangian
density constructed for $S^{2|2}$ we found that the field
equations for $\Phi$ reduce, when written in terms of the
component fields, to the ones derived directly from the action on
the ordinary sphere. The super conservation equations derived from
Noether's theorem --- when applied to the Lagrangian density for
the supersphere --- give rise to the familiar energy-momentum
tensor and fermionic currents expected from the component field
action. Notably, though, the super conservation equations also
give rise to additional conservation laws, that appear to be
independent of the familiar ones and which thus call for some
interpretation.

In this work we have concentrated on superscalar field theories on
the supersphere. Using the methods we have presented it would be
possible to further this study by investigating more general field
theories, for example gauge theories or sigma models with the
supersphere as the base space. Another possible extension of this
work would be to quantize the scalar field theory, which due to
its Euclidean nature would correspond to a statistical field
theory.

%\vspace{-2ex}

%% file: acknowledgements.tex
\section*{\normalsize{Acknowledgements}}
%\vspace{-2ex}
The authors would like to thank
Prof.\/~N.\/~S.\/~Manton for suggesting this project and for many
helpful conversations.

This work was partly supported by the UK Engineering and Physical Sciences
Research Council. A.F.S.\/~gratefully acknowledges financial support by
the Gates Cambridge Trust.

%% file: appendix.tex
\begin{appendix}
\section{Appendix}
\subsection{Raising and lowering conventions for spinor
  indices}\label{sec:epsilon}
Raising and lowering of spinor indices $\alpha,\beta,\ldots$ is
achieved with the use of the antisymmetric epsilon symbols
$\epsilon^{\alpha\beta}$ and $\epsilon_{\alpha\beta}$; the
convention we will follow is that of \cite{madness}. When raising
an index we always contract on the second index of
$\epsilon^{\alpha\beta}$, e.g.\/~
\be
\psi^\alpha=\epsilon^{\alpha\beta}\psi_\beta.
\ee
However, when lowering an index we always contract on the first index of
$\epsilon_{\alpha\beta}$, e.g.\/~
\be
\psi_\beta=\epsilon_{\gamma\beta}\psi^\gamma.
\ee
Combining the previous two equations we see that
\be
\epsilon^{\alpha\beta}\epsilon_{\gamma\beta}=\delta^\alpha{}_\gamma.
\ee
Hence we see that if we choose $\epsilon_{-+}=1$, then we must
also have $\epsilon^{-+}=1$. Note that we can think of
$\epsilon^{\alpha\beta}$ as $\epsilon_{\alpha\beta}$ with both
indices raised.

The (components of the) standard Pauli matrices are taken to be
$(\sigma^i)^\alpha{}_\beta$. Lowering the first index allows us to
construct the quantity
\be\label{eqn:epsilonsigma}
(\sigma^i)_{\alpha\beta} =
\epsilon_{\gamma\alpha}(\sigma^i)^\gamma{}_\beta =
-\epsilon_{\alpha\gamma}(\sigma^i)^\gamma{}_\beta,
\ee
which is symmetric in $\alpha, \beta$. We can then raise the
second index to give
\be\label{eqn:sigmatranspose}
(\sigma^i)_\alpha{}^\beta =
\epsilon_{\gamma\alpha}(\sigma^i)^\gamma{}_\delta
\epsilon^{\beta\delta} =
\epsilon_{\alpha\gamma}(\sigma^i)^\gamma{}_\delta
\epsilon^{\delta\beta}.
\ee
Notice that the third terms in
\refec{eqn:epsilonsigma}{eqn:sigmatranspose} have been written in
a way more suggestive of standard matrix multiplication. In fact,
if we define the antisymmetric matrix
$\epsilon=(\epsilon_{\alpha\beta})$, we may think of these
quantities as the components of the matrices $-\epsilon\sigma^i$
and $\epsilon\sigma^i\epsilon=(\sigma^i)^\transpose$ respectively.

The quantity $(\sigma^i)_\alpha{}^\beta$ (as well as
$(\sigma^a)_\alpha{}^\beta$) is a $\mathcal{U}(1)$ invariant
tensor, i.e.
\begin{equation}
(\sigma^i)_\alpha{}^\beta = R_\alpha{}^\gamma(h)
(\sigma^j)_\gamma{}^\delta R_\delta{}^\beta (h^{-1})
R_j{}^i(h^{-1}),
\end{equation}
where $R_p{}^q(h)$ is as given in \refe{eqn:coadjointrep}.

\subsection{Superzweibein and spin connection in polar coordinates}
\label{sec:appendixpolar} We obtain for the superzweibein in
(super)-polar coordinates
{\renewcommand{\arraystretch}{1.9}%
{\setlength{\arraycolsep}{0.8mm}%
\begin{equation} \label{eqn:superzweibeinpolar}(\tilde{E}_M{}^A) =
{\mbox{\scriptsize
$ \left(%
\! \begin{array}{cc|cc}
\sin\!\theta & 0 & 0 & 0 \\
0 & -1 & 0 & 0 \\ \hline \frac{i}{4}(\eta^\diamond \sin \! \theta
- \eta \cos\! \theta e^{- i \varphi} ) & - \frac{1}{4} \eta e^{- i
\varphi} & (1 + \frac{1}{8} \eta \eta^\diamond)
\sin\!\frac{\theta}{2} e^{-i \varphi/2} & (1 + \frac{1}{8} \eta
\eta^\diamond) \cos\!
\frac{\theta}{2} e^{- i \varphi/2} \\
\frac{i}{4}(\eta \sin\! \theta + \eta^\diamond \cos\! \theta e^{ i
\varphi}) & -\frac{1}{4} \eta^\diamond e^{i \varphi} & (1 +
\frac{1}{8} \eta \eta^\diamond) \cos\! \frac{\theta}{2} e^{i
\varphi/2} & - (1 + \frac{1}{8} \eta \eta^\diamond) \sin\!
\frac{\theta}{2} e^{i \varphi/2}
\end{array}%
\!\right)$}},
\end{equation}}}%
where the index $M$ here runs over $\varphi, \theta, \eta,
\eta^\diamond$. The spin connection is in polar coordinates given
by
\begin{align}
\omega^0 & = - d \varphi  \cos\!\theta - \frac{i}{4} d
\eta^\diamond (\eta \cos\!\theta - \eta^\diamond e^{i \varphi}
\sin\!\theta) - \frac{i}{4} d
\eta (\eta^\diamond \cos\!\theta + \eta e^{-i \varphi} \sin\!\theta) \\
\nonumber {} & = - \cot\!\theta \tilde{E}^1 + \frac{i}{4}
\frac{1}{\sin\!\theta} (\eta e^{-i \varphi/2}
\sin\!\frac{\theta}{2}
 - \eta^\diamond e^{i \varphi/2} \cos\!\frac{\theta}{2}) \tilde{E}^-  \\ \label{eqn:spinconnectionpolar}
 & {} \hspace{12.5ex} + \frac{i}{4} \frac{1}{\sin\!\theta}
 (\eta e^{-i \varphi/2} \cos\!\frac{\theta}{2}
 + \eta^\diamond e^{i \varphi/2} \sin\!\frac{\theta}{2}) \tilde{E}^+ .
\end{align}

\subsection{Superzweibein and spin connection in the $(w, \zeta)$ patch}
\label{sec:appendixwzeta} We find for the superzweibein in the
$(w, \zeta)$ coordinate patch
{\renewcommand{\arraystretch}{1.9}%
{\setlength{\arraycolsep}{0.6mm}%
\begin{equation}
\label{eqn:superzweibeinwzeta} (E'_M{}^A) =
\left(%
\! \begin{array}{cc|cc}
  \frac{i}{1 + w w^\diamond + \zeta \zeta^\diamond} & \frac{-1}{1 + w w^\diamond + \zeta \zeta^\diamond} &
  \frac{2i (\zeta w^\diamond + \zeta^\diamond)}{(1 + ww^\diamond)^{3/2}} & 0 \\
  \frac{- i}{1 + w w^\diamond + \zeta \zeta^\diamond} &  \frac{-1}{1 + w w^\diamond + \zeta \zeta^\diamond} & 0 &
  \frac{2i (\zeta^\diamond w - \zeta)}{(1 + ww^\diamond)^{3/2}}
  \\ \hline
  \frac{-i \zeta}{1 + ww^\diamond} &  \frac{\zeta}{1 + ww^\diamond} & \frac{-2i}{(1 + ww^\diamond - \zeta \zeta^\diamond)^{1/2}}  &  0 \\
  \frac{i \zeta^\diamond}{1 + ww^\diamond} & \frac{\zeta^\diamond}{1 + ww^\diamond} & 0 & \frac{-2i}{(1 + ww^\diamond - \zeta \zeta^\diamond)^{1/2}}  \\
\end{array}%
\!\right),
\end{equation}}}%
where the index $M$ now runs over $w, w^\diamond, \zeta,
\zeta^\diamond$. The inverse superzweibein is given by
\renewcommand{\arraystretch}{2}%
{\setlength{\arraycolsep}{0.5mm}%
\begin{equation}
\label{eqn:inversesuper2beinwzeta} (E'_A{}^M) = {\mbox{\scriptsize
$\left(%
\! \begin{array}{cc|cc}
  - \frac{i}{2}(1 + w w^\diamond) & \frac{i}{2}(1 + ww^\diamond) &
 - \frac{i}{2} (\zeta w^\diamond + \zeta^\diamond) & \frac{i}{2} (\zeta^\diamond w  -  \zeta)\\
  - \frac{1}{2}(1+  ww^\diamond) & - \frac{1}{2}(1 + ww^\diamond) & - \frac{1}{2} (\zeta w^\diamond + \zeta^\diamond)&
  - \frac{1}{2} (\zeta^\diamond w - \zeta)
  \\\hline
  \frac{i}{2}(1 + ww^\diamond)^{1/2}\zeta &  0 & \frac{i}{2}(1 + ww^\diamond + \zeta \zeta^\diamond)^{1/2}  &  0 \\
  0  & \frac{i}{2}(1 + ww^\diamond)^{1/2}\zeta^\diamond & 0 & \frac{i}{2}(1 + ww^\diamond+ \zeta \zeta^\diamond)^{1/2}  \\
\end{array}%
\! \right)$}}.
\end{equation}}
\renewcommand{\arraystretch}{1}%
The superdeterminant of $(E'_M{}^A)$ is given by
\begin{equation}
\label{eqn:sdetwzeta} E' \equiv \sdet(E'_M{}^A) = \frac{i}{2}\; \frac{1}{1
+ ww^\diamond + \zeta \zeta^\diamond}.
\end{equation}
Finally, we find for the spin connection in the $(w, \zeta)$ coordinate
patch
\begin{align}
\label{eqn:spinconnectionwzeta} \omega'^0 & = \frac{i}{1 +
ww^\diamond + \zeta \zeta^\diamond}(w^\diamond dw - w dw^\diamond
+ d\zeta \zeta^\diamond + d\zeta^\diamond \zeta)\\
\nonumber {} & = \frac{1}{2}(w^\diamond + w) E'^1 -
\frac{i}{2}(w^\diamond - w) E'^2
\\
& \hspace{2.5ex} + \frac{1}{2} \frac{\zeta w^\diamond +
\zeta^\diamond}{(1 + w w^\diamond)^{1/2}} E'^- - \frac{1}{2}
\frac{\zeta^\diamond w - \zeta}{(1 + w w^\diamond)^{1/2}} E'^+ .
\end{align}

\subsection{Euler-Lagrange equations for the full action}
\label{sec:appendixeqofmotion} The field equations following from
the full action given in \refe{eqn:fullaction} are
\begin{align}
\label{eqn:fieldeqfullaction1}
\partial_z \partial_{z^\diamond} A & =  \frac{1}{2} \frac{U'(A)}{(1 + z
z^\diamond)^2} - \frac{1}{2} \frac{F U''(A)}{1 + z z^\diamond} +
\frac{1}{2} \frac{\psi_\chi \psi_{\chi^\diamond} U'''(A)}{1 + z
z^\diamond} \\
\label{eqn:fieldeqfullaction2}
\partial_{z^\diamond} \psi_\chi & =  - \frac{1}{2}\frac{\psi_{\chi^\diamond}U''(A)}{1 + z
z^\diamond} \\
\label{eqn:fieldeqfullaction3}
 \partial_{z} \psi_{\chi^\diamond} & = \frac{1}{2}\frac{\psi_{\chi}U''(A)}{1 + z
z^\diamond} \\
\label{eqn:fieldeqfullaction4} F & = \frac{1}{2} \frac{U'(A)}{1 + z
z^\diamond}.
\end{align}

\subsection{Conformal invariance of the kinetic part of the action}
\label{sec:conformal}

The superscalar field action, \refe{eqn:fullaction}, can be
rewritten using the notation of Section \ref{sec:killing}. We find
it to be
\begin{align}
I & =\frac{i}{16}\int d^2z\;\sqrt{|g|} \bigg(
g^{mn}\partial_m\tilde
  A\partial_n\tilde A +
  i\boldsymbol{\psi}^\ddagger\partial\!\!\!/\boldsymbol{\psi}
  + \tilde F^2 \nonumber \\
  & \hspace{15ex}
  -\frac{1}{2}\boldsymbol{\psi}^\ddagger\boldsymbol{\psi}U''(\tilde
  A) + U(\tilde A) + \tilde F U'(\tilde A) \bigg),
\end{align}
where $g$ is the determinant of the metric. The kinetic part of the
action is obtained by setting $U(\tilde A)=0$. Note that we could replace the
second term, $i\boldsymbol{\psi}^\ddagger\partial\!\!\!/\boldsymbol{\psi}$,
with $i\boldsymbol{\psi}^\ddagger D\!\!\!\!/\boldsymbol{\psi}$. This
is because the term involving the spin connection will vanish due to
the anticommuting nature of $\boldsymbol{\psi}$ and the form of the
gamma matrices.

Under a conformal transformation, the metric and gamma matrices transform as
\begin{align}
g_{mn}&\to \lambda^2 g_{mn} \\
\gamma^m &\to \lambda^{-1}\gamma^m
\end{align}
where $\lambda$ is some positive function on the sphere. It is
then possible to define the transformation properties of the
component fields in such a way that the kinetic part of the action
will remain invariant. We find
\begin{align}
\tilde A &\to \tilde A, \\
\boldsymbol{\psi} &\to \lambda^{-\frac{1}{2}}\boldsymbol{\psi}, \\
\tilde F &\to \lambda^{-1} \tilde F.
\end{align}
The presence of a non-zero potential will break this conformal invariance.

\subsection{Transformation properties of superscalar
  densities}\label{sec:superscalardensity}

Using the infinitesimal point transformation $X'^M=X^M+\delta u \,
\Xi^M(X)$ we can define the Lie derivative of any supertensor
field $T(X)$ by
\be
£_\Xi T(X)=\lim_{\delta u \to 0}\frac{T(X')-T'(X')}{\delta u}.
\ee
For instance, a superscalar transforms as $\Phi'(X')=\Phi(X)$,
hence the Lie derivative can be calculated by using a Taylor
expansion. We find
\be
£_\Xi \Phi(X)=\Xi^M\partial_M \Phi(X).
\ee
Now, let $\mathfrak{T}(X)$ be a superscalar density of weight $+1$. It is
defined to transform as
\be
\mathfrak{T}'(X')=J(X)\mathfrak{T}(X),
\ee
where $J(X)$ is given by the superdeterminant
\begin{align}
J(X) & = \sdet \left(\frac{\partial X^M}{\partial X'^N}\right) \\
& = 1- \delta u (-1)^M\partial_M \Xi^M + \ldots
\end{align}
Note that in the last line we have expanded the superdeterminant to first
order, resulting in the appearance of a supertrace, this explains the
factor $(-1)^M$ in the summation over $M$. Also we can
expand
\be
\mathfrak{T}(X')=\mathfrak{T}(X)+\delta u \,\Xi^M\partial_M
\mathfrak{T}(X)+\ldots
\ee
Combining these gives us the Lie derivative of a superscalar density
\be
£_\Xi \mathfrak{T}(X) = (-1)^M \partial_M (\Xi^M \mathfrak{T}(X)).
\ee
The same procedure can be used to calculate the Lie derivative of any
supertensor field.

Using the Lie derivative we can describe the infinitesimal active
coordinate transformation, $X\to X+\delta u \,\Xi$, alternatively
as a transformation of the fields. We need to find the difference
between the tensor which has been dragged along $\delta u \,\Xi$
to the point $X$, and the tensor which was already at $X$. For the
supertensor field $T(X)$ this difference is given by
\be
\delta T(X)=-\delta u £_{\Xi}T(X).
\ee

\end{appendix}

%% file: references.tex
\providecommand{\href}[2]{#2}\begingroup\raggedright\endgroup